\documentclass[superscriptaddress,amsmath,amssymb,aps,pra,onecolumn]{revtex4-2}
\usepackage{subfigure}
\usepackage{graphicx}
\usepackage{dcolumn}
\usepackage{bm}
\usepackage[normalem]{ulem}
\usepackage{amssymb}
\usepackage{soul}
\usepackage{xcolor}
\usepackage{xparse}
\usepackage{color}
\usepackage{comment}
\usepackage{pgfplots}
\pgfplotsset{compat=1.9}
\usepackage{tikz}
\usepackage{hyperref}

\usepackage{amsmath}
\DeclareMathOperator{\sech}{sech}
\DeclareMathOperator{\csch}{csch}
\DeclareMathOperator{\spanOperator}{span}
\DeclareMathOperator{\argOperator}{arg}
\DeclareMathOperator{\Trace}{Tr}
\DeclareMathOperator{\Real}{Re}
\DeclareMathOperator{\Imaginary}{Im}

\usepackage[normalem]{ulem}

\begin{document}

\title{Bright Solitons in Non-Hermitian systems: \\
Exploring the interplay of $\mathcal{PT}$ Symmetry and Wadati Structure}

\author{Sathyanarayanan Chandramouli}
\affiliation{Okinawa Institute of Science and Technology, Onna, Okinawa 904-0497, Japan}
\affiliation{Department of Mathematics and Statistics, University of Massachusetts at Amherst}

\author{R. Carretero-Gonz\'alez}
\affiliation{Nonlinear Dynamical Systems Group\footnote{{\tt URL:} http://nlds.sdsu.edu/},
Computational Science Research Center\footnote{{\tt URL:} http://www.csrc.sdsu.edu/}, Department of Mathematics and Statistics, San Diego State University}

\author{D.J. Frantzeskakis}
\affiliation{Department of Physics, National and Kapodistrian University of Athens, Greece}

\author{P.G. Kevrekidis}
\affiliation{Department of Mathematics and Statistics, University of Massachusetts at Amherst}

\date{\today}

\begin{abstract}
In this study, we revisit the subject of optical solitons in non-Hermitian potentials. For the underlying model of the focusing nonlinear Schrödinger equation with a complex localized potential, we offer a classification of such models
into four canonical cases. These correspond to 
(a) $\mathcal{PT}$-symmetric Wadati potentials, 
(b) $\mathcal{PT}$-symmetric non-Wadati potentials, 
(c) Non-$\mathcal{PT}$ but Wadati potentials, and 
(d) complex potentials with neither $\mathcal{PT}$ nor Wadati symmetries. 
We study the existence and stability of optical solitons across prototypical model variations spanning all four relevant quadrants.
We complement such an investigation with a low-dimensional effective description based on a non-conservative variational approximation method. The variational framework is assessed in terms of its ability to represent the structural properties and bifurcations of solutions. {Our results suggest that continuous families of bright solitons persist across sectors (a)--(c). 
For non-$\mathcal{PT}$-symmetric potentials, small departures from the Wadati limit, corresponding to sector (d), suggest the loss of continuous families of bright-solitons including, more specifically, the possibility of the generic absence of single-humped localized states. 
The corresponding variational analysis is broadly consistent with these observations.
We also highlight the usefulness and limitations for such variational approaches
in describing bright solitons in the presence of non-Hermitian potentials.} 
\end{abstract}

\maketitle


\section{Introduction}
\label{sec:introduction}

The notion of $\mathcal{PT}$-symmetry lies at the heart of non-Hermitian physics~\cite{ashida2020non,el2018non,konotop2016nonlinear,ChristodoulidesYang2018}, which seeks to describe non-conservative systems that interact with their external environment. Historically, the subject of non-Hermitian quantum physics attracted considerable interest following the seminal observation that certain non-Hermitian Hamiltonians possessing parity ($\mathcal{P}$) and time-reversal ($\mathcal{T}$) symmetry can exhibit an entirely real eigenspectrum~\cite{bender1998real}. Remarkably, this demonstrated that Hermiticity, or equivalently the self-adjointness of the associated Schrödinger operator, is not a necessary condition for the corresponding  spectrum to be real. 
These developments subsequently motivated the study of broader classes of non-Hermitian operators, including pseudo-Hermitian operators admitting purely real spectra~\cite{mostafazadeh2001pseudo,mostafazadeh2002pseudo}.

The structural equivalence between the single-particle Schrödinger equation and the paraxial wave equation in classical optics opened the door to the rapidly growing field of $\mathcal{PT}$-symmetric optics~\cite{FengElGanainyGe2017}. By mapping the quantum mechanical potential to a complex-valued refractive index profile, where the real part represents the waveguide's refractive index and the imaginary part represents balanced gain and loss, this paradigm has sparked intense research into non-Hermitian optical beam propagation and the parallels between optics and quantum mechanics at both the linear and the nonlinear level~\cite{makris2008beam,el2007theory,musslimani2008optical,makris2010pt,ruschhaupt2005physical}.
Substantial experimental progress has since been made across a wide range of photonic platforms for realizing and probing $\mathcal{PT}$-symmetric effects, including optical waveguides~\cite{guo2009observation,ruter2010observation}, photonic lattices~\cite{regensburger2012parity}, optical metamaterials~\cite{feng2013experimental}, optical microresonators~\cite{chang2014parity}, and coherently prepared warm atomic gases~\cite{zhang2016observation,hang2017parity}.
Furthermore, the notion of non-Hermitian wave propagation was explored in various other disciplines, e.g., non-Hermitian acoustics~\cite{huang2024acoustic}, $\mathcal{PT}$-symmetric metamaterials~\cite{pr2025unravelling}, exciton-polaritons and plasmonics~\cite{gao2015observation}, ultracold Fermi gases~\cite{li2019observation}, and in electrical circuits~\cite{schindler2011experimental} just to name a few; see Refs.~\cite{konotop2016nonlinear,ashida2020non} for comprehensive reviews. 
%
%
Wadati introduced a systematic construction of $\mathcal{PT}$-symmetric complex potentials by exploiting the correspondence between Dirac- and Schrödinger-type spectral problems arising in inverse-scattering theory~\cite{wadati2008construction}.
Subsequent work showed that the same structural class can remain spectrally real and support continuous nonlinear-mode families even without $\mathcal{PT}$ symmetry~\cite{tsoy2014stable,konotop2014families}.
Since then, the so-called Wadati potentials have by themselves opened several avenues of investigation~\cite{barashenkov2016exactly,zezyulin2022nonlinear,yang2021analytical,bagchi2020new,makris2015constant,makris2017wave}. A notable property of the nonlinear Schrödinger (NLS) equation with a non-Hermitian Wadati potential concerns the existence of continuous families of stable solitons for {non}-$\mathcal{PT}$ asymmetric 
potentials~\cite{yang2021analytical,konotop2014families}. Conversely, such systems can also support continuous families of asymmetric solitons within strictly $\mathcal{PT}$-symmetric landscapes~\cite{yang2014symmetry}, as well as the existence of constant-intensity waves~\cite{makris2015constant,makris2016constant}. Furthermore, hierarchies of integrable partial differential equations (PDEs) have been established by extending these non-Hermitian principles to nonlocal, $\mathcal{PT}$-symmetric variants of the NLS equation~\cite{ablowitz2016inverse}.


Self-localized states (solitons) in purely non-Hermitian settings typically exist as isolated solutions rather than as continuous families parameterized by a propagation constant~\cite{Yang2014Necessity,zezyulin2022nonexistence}. However, with additional symmetries
---e.g., $\mathcal{PT}$-symmetry, i.e., balanced gain-loss distributions or underlying integrals of motion which exist for so called ``Wadati" potentials---
such continuous families could still be found (cf.~Refs.~\cite{wimmer2015observation,konotop2014families}). The existence and stability of such stationary modes are central to understanding various dynamical phenomena (such as scattering, cf.~Refs.~\cite{al2013unidirectional,abdullaev2014scattering}) in these non-Hermitian systems. Despite extensive previous studies addressing these questions in non-Hermitian systems possessing (or lacking) one or more of these properties, comparatively few studies have systematically examined soliton existence and stability across all possible combinations of $\mathcal{PT}$ symmetry and Wadati structure. In this work, we address this question within a unified framework by considering the four ``quadrants'' of a classification of hon-Hermitian systems that we introduce
based on the presence or absence of these two properties:
\begin{itemize}
\itemsep 0pt
\item[(a)] systems possessing both $\mathcal{PT}$ symmetry and Wadati structure,
\item[(b)] systems possessing $\mathcal{PT}$ symmetry but lacking Wadati structure,
\item[(c)] systems possessing Wadati structure but lacking $\mathcal{PT}$ symmetry, and
\item[(d)] systems possessing neither $\mathcal{PT}$ symmetry nor Wadati structure.
\end{itemize}
Furthermore, we examine the efficacy of low-dimensional descriptions within each of these cases in predicting soliton existence and stability.
Quadrants (a)--(c) are found to possess continuous (in a canonical parameter,
such as the solution frequency) families of standing bright solitary waves.
The variational formulation begins to exhibit qualitative discrepancies 
in sector (c), most notably at the level of the linearization spectrum, 
underscoring the richness of the Wadati structure and the challenges 
associated with capturing it within a low-dimensional variational framework. 
We are unable to identify continous such families in sector (d), where the
isolated solutions we identify, in fact, also no longer preserve a single-humped
spatial profile, in line with other \textit{generic} non-Hermitian systems such as the complex potentials of Ref.~\cite{zezyulin2022nonexistence}.

Our presentation is organized as follows. In Sec.~\ref{Section-2}, we present the NLS model with a complex external potential considered in this work. In Sec.~\ref{NCVA-formulation}, we develop a low-dimensional description of its bright soliton solutions using a non-conservative variational approximation. In Sec.~\ref{Results}, we discuss the existence, stability, and dynamical properties of bright solitons across the four quadrants introduced above.
Finally, Sec.~\ref{Conclusions} summarizes the main findings and outlines 
some outstanding challenges and directions for future work.

\section{The nonlinear Schrödinger equation with complex potential }
\label{Section-2}

In the paraxial approximation, the evolution of optical beams in media featuring 
a focusing cubic (Kerr) nonlinearity and a spatially dependent refractive index 
and a gain-loss distribution is governed by the following dimensionless NLS 
equation~\cite{ChristodoulidesYang2018,kivshar_optical_2003}:
%
%
\begin{equation}
\label{NLS-nonHermitian-potential}
i \Psi_t + \frac{1}{2} \Psi_{xx} 
+ \left[V_R(x)+i\,V_I(x)\right] \Psi +|\Psi|^2 \Psi=0,
\end{equation}
where the $t$ and $x$ subscripts denote partial derivatives. 
{Here, the wave function $\Psi(x,t)$ represents the complex electric field envelope, the variables $t$ and $x$ denote the propagation and transverse directions, respectively, while}
$V_{R}$ and $V_{I}$ are real-valued functions representing the  
refractive index and the gain-loss distribution,  respectively. In terms of 
$\Psi(x,t)$, the beam intensity $|\Psi|^2$ and the transverse power flow density $\frac{1}{2i}(\Psi^{*}\Psi_x-\Psi_x^{*}\Psi)$ (asterisks denote complex conjugation) are generically not conserved in the presence of the imaginary part of the potential. 
In this work, we restrict ourselves to the scenario {where $\Psi$ evolves in the presence of a} spatially localized inhomogeneity, i.e., when 
{both $V_R(x)$ and $V_I(x)$ decay} exponentially fast.
The case of periodic potentials, e.g., in the spirit of Refs.~\cite{musslimani2008optical,makris2010pt}
is interesting in its own right as a subject of future study.

{Stationary} solitons or nonlinear eigenmodes, of the form $\Psi(x,t)=\Phi_0(x) e^{i\mu t}$, where $\mu$ is the propagation constant (or frequency), satisfy the following boundary value problem
\begin{equation}
\label{nonlinear-eigenmodes}
    -\mu \Phi_0+\frac{1}{2}\Phi_0''+V\Phi_0+|\Phi_0|^2\Phi_0=0,
\end{equation}
where the complex wavefunction $\Phi_0(x)$ satisfies the boundary 
conditions $|\Phi_0(\pm\infty)|=0$.
We first review the properties of solitons along the two axes of $\mathcal{PT}$ symmetry and Wadati structure, respectively. $\mathcal{PT}$ symmetry entails invariance under the combined actions of parity ($\mathcal{P}$) and time reversal ($\mathcal{T}$), with the latter involving complex conjugation, i.e.,
$\Psi(x,t)\xrightarrow{\mathcal{PT}}\Psi^*(-x,-t)$, which is permissible generally for $V_R(-x)=V_R(x)$ (even) and $V_I(-x)=-V_I(x)$ (odd). Notably, the odd parity of the imaginary part $V_I(x)$ entails a ``global" balance of gain and loss (between the left and right sides), rendering possible the existence of continuous families of stationary modes~\cite{konotop2016nonlinear,konotop2014families}.
For $\mathcal{PT}$-symmetric stationary modes 
with real $\mu$, this symmetry is inherited by the stationary profile according to
$\Phi_0(x)=\Phi_0^*(-x)$, i.e., the real (imaginary) parts possess even (odd) parity. 
Typically, the $\mathcal{PT}$-symmetry of the model encodes the same symmetry onto the soliton, except  in the case of certain particular potentials; see, for instance, Ref.~\cite{yang2014symmetry}.

On another note, the family of Wadati potentials was first introduced by Wadati~\cite{wadati2008construction} in the context of constructing $\mathcal{PT}$-symmetric potentials for the Schrödinger operator using inverse scattering theory. The NLS with a Wadati potential possesses special properties that are not necessarily shared by the broader class of $\mathcal{PT}$-symmetric potentials~\cite{barashenkov2016exactly}. In particular, Wadati potentials need not satisfy the $\mathcal{PT}$-symmetry condition. Instead, the real and imaginary parts of the potential are defined in terms of the function $\mathcal{W}(x)$ as 
$$
    V_R(x)=\frac{1}{2}\mathcal{W}^2,
    \quad
    V_I(x)=-\frac{1}{2}\mathcal{W}_x.
$$
A crucial structural property of Wadati potentials is the existence of a first integral for the stationary spatial-dynamics system, which removes one of the independent solvability constraints and thereby permits continuous families of localized modes~\cite{NixonYang2016,nixon2016nonlinear}. 
Another important property of Wadati systems is the existence of constant-intensity waves~\cite{makris2015constant}, as well as the transformation $\Psi=\psi\exp(i\int_x \mathcal{W}(x') dx')$, which renders Eq.~\eqref{NLS-nonHermitian-potential} into the following inhomogeneous, advective model:
\begin{equation}
\label{Inhomog-advective}
    i\psi_t+i\mathcal{W}(x)\psi_x+\frac{1}{2}\psi_{xx}+|\psi|^2\psi=0.
\end{equation}

In this work, we explore the four cases defined by the axes of $\mathcal{PT}$ symmetry and Wadati structure. Representative exponentially localized potentials considered in each quadrant are
\begin{itemize}
    \item[(a)] $\mathcal{PT}$-symmetric Wadati potential defined by the even-parity potential function 
    \begin{equation}
    \label{PT-Wadati-potential-function}
       \mathcal{W}(x)=\alpha \sech^2(x),\;\;{\rm with}\;\alpha>0. 
    \end{equation} 
    
             \item[(b)] $\mathcal{PT}$-symmetric potentials lacking Wadati structure. Here, we consider a family of potentials with
\begin{equation}
       \label{PT-symmetry-but-no-Wadati}
        V_R(x)=\frac{\alpha^2}{2}\sech^4(x), \quad V_I(x)= \frac{3}{2}\alpha\tanh(x)\sech^2(x).
    \end{equation}
    
        \item[(c)] Wadati potentials lacking $\mathcal{PT}$-symmetry. Here, the potential $V$ is defined by the potential function $\mathcal{W}(x)$. 
        {In this case,} we consider the family of asymmetric single-hump potentials
\begin{equation}
    \label{WADATI-no-PT}
       \mathcal{W}(x)=\alpha\sech^2(x)+\Gamma\alpha\sech^2(x-x_0),
       \end{equation}
       {and} we choose a weak degree of asymmetry in the potential function, i.e., $\Gamma\ll1$ ($\Gamma>0$), and $0<x_0\sim 1$.

    \item[(d)] Potentials lacking both the $\mathcal{PT}$-symmetry and Wadati structure. Here, we consider a family of potentials 
    \begin{subequations}
    \label{VR_VI-non-conservative}
    \begin{eqnarray}
    \label{VR-non-conservative}
        V_R(x)&=&\frac{1}{2}\left[\alpha\sech^2(x)+\Gamma\alpha\sech^2(x-x_0)\right]^2, \\
    \label{VI-non-conservative}
        V_I(x)&=&\frac{1-\beta}{2}\left[2\alpha\tanh(x)\sech^2(x)
        +2\Gamma\alpha\tanh(x-x_0)\sech^2(x-x_0)\right],
    \end{eqnarray}
    \end{subequations}
   wherein the potentials are perturbed away from an appropriate Wadati potential by tuning the proximity parameter $\beta\sim\Gamma\ll 1$. 
    It is worth stressing again that, generically, such a non-$\mathcal{PT}$ and non-Wadati system {is expected to} only have solutions at \textit{isolated} values of $\mu$~\cite{Yang2014Necessity,zezyulin2022nonexistence}.
\end{itemize}

We study the existence and stability of bright modes bifurcating from the bulk NLS limit ($\alpha=0$). We employ variational approximations to derive a reduced-order description of the bright modes. Such approximations have been successfully applied in a variety of related contexts to non-conservative NLS models~\cite{rossi2020non,rossi2024temporal}.
In particular, we seek to assess the feasibility of a low-dimensional representation of the bright modes in capturing their existence, stability, and bifurcation structure.
%

\section{Non-conservative variational approximation}
\label{NCVA-formulation}

We utilize a low-dimensional description of the bright soliton eigenmodes 
to the NLS with non-Hermitian potential.
To this end, we reduce the {NLS} dynamics 
to an effective “particle” model, {i.e.,}
to a solitonic manifold described by a set of coupled ODEs governing the evolution of the bright-soliton parameters (amplitude, position, width, phase, velocity, and chirp). The underlying solitonic manifold construction is based on the non-conservative variational approximation (NCVA) for Schrödinger-type systems, as outlined in Refs.~\cite{rossi2020non,rossi2024temporal}. The method draws inspiration from the Lagrangian (and Hamiltonian) formulation of non-conservative mechanical systems, as developed in Refs.~\cite{galley_classical_2013,galley2014principle}. 
Inspired by these {earlier} works, the extension to non-conservative field theories (i.e., PDEs) was studied in Ref.~\cite{kevrekidis2014variational} with the $\mathcal{PT}$-symmetric variants of the sine-Gordon and $\phi^4$-models. 

We set the stage by factoring out the linear phase accumulation associated with soliton propagation, through the substitution $\Psi(x,t)=\Phi(x,t)e^{i\mu t}$,
yielding
\begin{equation}
\label{Phi-mu}
i\Phi_t-\mu\Phi+V(x)\Phi+\frac{1}{2}\Phi_{xx}+|\Phi|^2\Phi=0.
\end{equation}
Furthermore, the explicit contribution of the real (conservative) part of the 
potential $V_R(x)$ can be eliminated through a spatially dependent phase 
transformation. To this end, we introduce the auxiliary function 
$\mathcal{W}(x)$ such that $\mathcal{W}^2/2=V_R$ and define 
$\Phi=\phi\, e^{i\theta(x)}$, with $\mathcal{W}(x)=\theta'(x)$, to obtain 
the following perturbed NLS equation:
\begin{equation}
\label{phi-time}
    i\phi_t-\mu\phi+\frac{1}{2}\phi_{xx}+|\phi|^2\phi
    =\mathcal{R}(\phi).
\end{equation}
Here, all non-conservative contributions $\mathcal{R}(\phi)=-i\mathcal{W}\phi_x-i\left(V_I+\mathcal{W}_x/2\right)\phi$ have been collected on the 
right-hand side (RHS). Notably, for a Wadati potential, 
$V_I=-\mathcal{W}_x/2$, leading to the cancellation of the second term on the RHS:
$V_I+\mathcal{W}_x/2\equiv 0$.

Hereafter, we consider a six-parameter solitonic-ansatz for such bright modes:
$$\overline{\phi}(x,\vec{p}(t)) = a \sech[w(x-\xi)]\exp[i(b(x-\xi)^2+c(x-\xi)+\chi)],$$ 
with $\vec{p}(t)=[a(t),b(t),c(t),\xi(t),w(t),\chi(t)]^T$, where $a(t)$, $b(t)$, $c(t)$, $\xi(t)$, $w(t)$, and $\chi(t)$ are the soliton amplitude, chirp, velocity, position, width, and phase (time-dependent) parameters. The formulation incorporates a chirp $b$, together with  its variational conjugate, the {(inverse)} soliton width parameter $w$. This pair of parameters is necessary to capture potential breathing dynamics~\cite{rossi2020non}, and to allow the soliton localization scale (set by $w$) to adjust to, and potentially be dictated by, the underlying complex potential. Focusing on the conservative components of the potential, these define the Lagrangian density
$$
\mathcal{L}_{\rm cons}=\frac{i}{2}(\phi \phi^{*}_t-\phi^{*}\phi_t)+\frac{1}{2}|\phi_x|^2-\frac{1}{2}|\phi|^4+\mu |\phi|^2,
$$
with an effective (averaged) Lagrangian given by
$$
     \overline{L}_{}=\int_{-\infty}^{\infty} \mathcal{L}_{\rm cons}\left[\overline{\phi}(x,\vec{p}(t))\right]dx=\frac{\pi^2 a^2 \dot {b}}{6 w^3}+\frac{a^2 w}{3}+\frac{\pi^2a^2b^2}{3 w^3}+\frac{c^2 a^2 }{w}-\frac{2 a^4}{3 w}+\frac{2\mu a^2}{w}+\frac{2 a^2}{w}(\dot{\chi}-c\dot{\xi}).
$$
Non-conservative variational calculus~\cite{rossi2020non} yields the following 
effective, Euler-Lagrange (EL), evolution equations:
$$
{\frac{d}{dt}\left( \frac{\partial \overline{L}}{\partial \dot{\vec{p}}} \right)-\frac{\partial \overline{L}}{\partial \vec{p}} -  I_{\vec{p}} = 0,
}
$$
wherein the overlapping integrals are given by
$$
I_{\vec{p}}=2\Real\left[\int_{-\infty}^{\infty}\mathcal{R}(\overline{\phi})\frac{\partial \overline{\phi}^*}{\partial \vec{p}}dx\right].
$$
These expressions arise from the non-conservative terms in the RHS of Eq.~\eqref{phi-time}, and are computed numerically using trapezoidal quadrature {(cf.~the Appendix~\ref{NCVA-Appendix} for details on their derivation)}.

The ODE system obtained from these EL equations can be explicitly represented in terms of the 
solitonic variables, 
{in the form} $\dot{\vec{p}}=\vec{F}(\vec{p})$, as follows:
\begin{eqnarray}
    \nonumber
    \dot a &=& -ab +\frac{3w}{8 a} I_{\chi}-\frac{3w^3}{2a\pi^2} I_b\\
    \nonumber
    \dot b &=& -2 b^2-\frac{2 a^2 w^2}{\pi^2}+\frac{2 w^4}{\pi^2}+\frac{3 w^4}{\pi^2 a^2}I_w+\frac{3w^3}{2\pi^2 a}I_a \\
    \nonumber
    \dot c &=&\frac{w}{2a^2}(-c I_{\chi}-I_{\xi}) \\
    \label{NCVA-ODE-system}
    \dot \xi &=& c+\frac{w}{2 a^2} I_c \\
    \nonumber
    \dot w &=&-2 b w+\frac{w^2}{4 a^2} I_{\chi}-\frac{3w^4}{a^2 \pi^2}I_b \\
    \nonumber
    \dot \chi &=& -\frac{3w}{8a}I_a-\frac{w^2}{3}+\frac{c^2}{2}+\frac{5 a^2}{6}+\frac{wc}{2a^2}I_c-\mu-\frac{w^2}{4 a^2}I_w.
\end{eqnarray}
%
{The fixed points $\vec{p}_0$ of this ODE system, 
satisfying $\vec{F}(\vec{p}_0)=\vec{0}$,} 
represent reduced order representations (approximations) of stationary 
bright soliton solutions to Eq.~\eqref{phi-time}. It is of interest to 
check if the bright solitonic ansatz captures the infinite-dimensional 
profile over a wide regime in parameter space.

It is useful to note that 
$\vec{p}_0$ may be 
recast by combining the conditions $F_a=0$ and $F_w=0$, which yield $I_{\chi}=0$.
In particular, this condition provides the NCVA counterpart of the PDE power-balance constraint $\int_{\mathbb{R}} |\Phi_0(x)|^2 V_I(x)\,dx=0$, and thereby selects the NCVA soliton center $\xi_0$.
Therefore, the six fixed-point equations for $\vec{p}_0$ 
can be written as
\begin{align}
\label{NCVA-fp-eqn-system-1}
I_{\chi} &= 0,\\
\label{NCVA-fp-eqn-system-2}
-2 b_0^2-\frac{2 a_0^2 w_0^2}{\pi^2}+\frac{2 w_0^4}{\pi^2}
+\frac{3 w_0^4}{\pi^2 a_0^2}I_w
+\frac{3w_0^3}{2\pi^2 a_0}I_a &= 0,\\
\label{NCVA-fp-eqn-system-3}
I_{\xi} &= 0,\\
\label{NCVA-fp-eqn-system-4}
c_0+\frac{w_0}{2a_0^2}I_c &= 0,\\
\label{NCVA-fp-eqn-system-5}
b_0+\frac{3w_0^3}{2a_0^2\pi^2}I_b &= 0,\\
\label{NCVA-fp-eqn-system-6}
-\frac{3w_0}{8a_0}I_a-\frac{w_0^2}{3}+\frac{c_0^2}{2}
+\frac{5a_0^2}{6}+\frac{w_0c_0}{2a_0^2}I_c
-\mu-\frac{w_0^2}{4a_0^2}I_w &= 0.
\end{align}
%
%
Upon fixing the arbitrary global phase, e.g., by setting $\chi_0\equiv0$, the system comprises six fixed-point equations for the five unknowns $\{a_0,b_0,c_0,\xi_0,w_0\}$. Thus, for a generic non-Hermitian potential, the reduced stationary problem is {\em overdetermined}. 
For the phase-transformed NLS with a Wadati potential, however, we have 
$I_{\xi}=2\Real[\int_{-\infty}^{\infty}i\mathcal{W}(x)|\partial_x\overline{\phi}(x,\vec{p})|^2\,dx]\equiv0$
identically on the variational manifold. Consequently, the condition $I_{\xi}=0$ imposes no additional constraint, leaving five independent fixed-point conditions for the five (nontrivial) variational parameters. Similarly, for a generic $\mathcal{PT}$-symmetric potential, a $\mathcal{PT}$-symmetric fixed point is characterized by $\xi_0=0$ (i.e., it is centered). Parity considerations then yield $I_{\chi}=0$ and $I_b=0$, with the latter, through the corresponding fixed-point equation, implying $b_0=0$. In turn, the conditions $\xi_0=b_0=0$ ensure that $I_{\xi}=0$ due to the underlying parity. Consequently, three of the six fixed-point conditions are satisfied identically, while the remaining three constitute independent constraints for the three nontrivial variational parameters $\{a_0,c_0,w_0\}$. Note that for generic $\mathcal{PT}$-symmetric potentials, symmetry breaking of $\mathcal{PT}$-symmetric solitons is anyway forbidden~\cite{yang2014can} with the exception of the Wadati potentials~\cite{yang2014symmetry}.

Besides existence, the linear stability of these fixed points (and hopefully, 
by extension, that of the bright solitons of the original NLS equation) can 
be probed by computing the eigenvalues $\{\lambda^{(j)}\}_{j=1}^{6}$ 
(and associated eigenvectors $\{\vec{v}^{(j)}\}_{j=1}^{6})$ of the Jacobian matrix $(d\vec{F}/d\vec{p})|_{\vec{p}_0}$. To connect the NCVA-ODE eigenmodes to their PDE {analogues}, we consider perturbing on the solitonic manifold along the $j$th NCVA eigendirection $\vec{v}^{(j)}$, namely:
$$
    \overline{\phi}(x,\vec{p}_0+\epsilon \vec{v}^{(j)})\approx \overline{\phi}(x,\vec{p}_0)+\epsilon \sum_{i=1}^{6} v^{(j)}_i\partial_{p_i} \overline{\phi}(x,\vec{p}_0), 
$$
where $\spanOperator\{\partial_{p_i} \overline{\phi}(x,\vec{p}_0)\}_{i=1}^{6}$ represents the local tangent space of the variational solitonic manifold at $\vec{p}_0$, $\vec{v}^{(j)}\in \mathbb{C}^6$
and an approximate PDE eigenmode is given by $\sum_{i=1}^{6} v^{(j)}_i\partial_{p_i} \overline{\phi}(x,\vec{p}_0)$. There are two aspects of interest when it comes to {the} fidelity of the NCVA stability picture: (i) Is the NCVA manifold tangent space $\spanOperator\{\partial_{p_i} \overline{\phi}(x,\vec{p}_0)\}_{i=1}^{6}$ ``rich" enough to represent the PDE eigenmodes? and (ii) does each of the eigendirections $\{\vec{v}^{(j)}\}$ combine the basis elements $\partial_{p_i} \overline{\phi}(x,\vec{p}_0)$ appropriately so as to represent the PDE eigenmode? To partially address (i), we note that the NCVA can, at best, capture the {\em discrete}, spatially localized PDE eigenmodes represented in the six-parameter ansatz, whereas the continuous-spectrum modes (that satisfy radiation boundary conditions), cannot be represented within the present finite-dimensional variational framework. In what follows, we will probe both of the above aspects for each of the four $\mathcal{PT}$-Wadati quadrant sectors outlined in the Introduction.

\begin{figure}
    \centering
    \includegraphics[width=\linewidth]{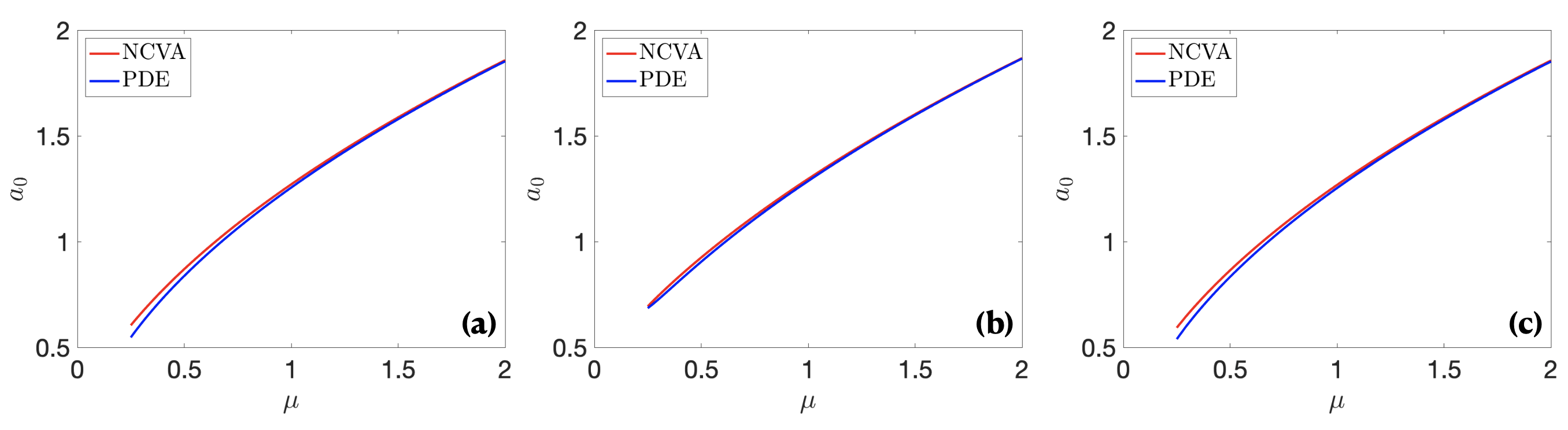}
    \caption{{The panels illustrate continuous families of solitary waves as characterized by the soliton amplitude as a function of the soliton frequency $\mu$ for: (a) the $\mathcal{PT}$-symmetric Wadati potential function [see Eq.~\eqref{PT-Wadati-potential-function}] with $\alpha=1$,
    (b) the $\mathcal{PT}$-symmetric potential [see Eq.~\eqref{PT-symmetry-but-no-Wadati}] with $\alpha=1$, and 
    (c) the non-$\mathcal{PT}$-symmetric Wadati potential function [see Eq.~\eqref{WADATI-no-PT}]. The other system parameters are fixed to $\alpha=1$, $x_0=2.75$, and $\Gamma=0.1$.  }
    }
    \label{fig:1-1}
\end{figure}

\section{Results}
\label{Results}
\subsection{Wadati potentials with $\mathcal{PT}$-symmetry}
\label{Wadati-PT-case}

Here, we consider a one-parameter family of $\mathcal{PT}$-symmetric 
Wadati potentials $V(x)$ as defined in Eq.~\eqref{PT-Wadati-potential-function}. 
Furthermore, we specifically seek $\mathcal{PT}$-symmetric stationary solutions 
to such a family of potentials, i.e., solutions satisfying $\Phi_0^*(-x)=\Phi_0(x)$. {Assuming $\mathcal{PT}$-symmetry removes the generic overdetermination of the stationary problem (as discussed in Secs.~\ref{Section-2} and \ref{NCVA-formulation}) and thereby permits continuous families of localized modes parametrized by $\mu$; see Fig.~\ref{fig:1-1}(a). 
In the present work, we display representative members of this families for $\mu=0.5$, for which the zero-potential limit corresponds to the unit-amplitude, unit-width NLS bright soliton. Our emphasis is therefore on the structural mechanisms governing stability, rather than on a systematic continuation of the solution families in $\mu$. An extension of the discussion can be made straighforwardly by introducing the rescaled coordinate $y=\sqrt{2\mu}\,x$ together with $\phi_0(x)=\sqrt{2\mu}\,u_0(y)$ for stationary solutions in Eq.~\eqref{phi-time} allowing for the propagation constant $\mu$ to be normalized, at the expense of a simultaneous rescaling of the strength \textit{and} spatial scale of the complex potential. Thus, while the quantitative soliton profiles and spectra generally depend on $\mu$, the structural mechanisms leading to instability investigated here are not specific to the specific selection of
the parameter $\mu$.}
Furthermore, we note that for the class of Wadati potentials considered here, with the associated function $\mathcal{W}(x)$ possessing a single extremum, the corresponding linear Schrödinger operator remains in the unbroken $\mathcal{PT}$-symmetric phase~\cite{tsoy2014stable}.

At the PDE level, we compute bright modes of the driven-dissipative system for $0<\alpha\lesssim 1.74$ using numerical continuation based on the Newton conjugate-gradient method~\cite{yang_newton-conjugate-gradient_2009}. The (approximate) upper endpoint, $\alpha\approx 1.74$, coincides with the linear limit, where the nonlinear bright mode bifurcates from an eigenmode of the underlying linear Schr\"odinger operator $\mathcal{L}_{\rm lin}=\frac{1}{2}\frac{d^2}{dx^2}+V(x)$. 
This algorithm can be initialized with the bright soliton solution of the NLS equation at $\alpha=0$, given by the profile ${\Phi}_0(x)=\sech(x)$, and then solutions are continued as one increases the value of the potential shape parameter $\alpha$. We store the corresponding PDE bright solitons in steps of $\Delta \alpha=0.05$.

Having computed this family of PDE modes, we analyze their linear stability by introducing perturbations to their profiles; that is $\Phi(x,t)=\Phi_0(x)+\epsilon(x,t)$ with $|\epsilon|\ll 1$. Substituting this, together with the expression for the complex-valued perturbation $\epsilon=\phi^I(x,t)+i\phi^R(x,t)$ into Eq.~\eqref{Phi-mu}, and collecting real and imaginary parts to leading order, yields the evolution equations governing the small-amplitude perturbations 
\begin{equation}
\label{Stability-bright-modes}
    \partial_t\begin{pmatrix}
      \phi^I\\\phi^R
    \end{pmatrix}= 
       \mathcal{L}
    \begin{pmatrix}
        \phi^I\\\phi^R
    \end{pmatrix},
\end{equation}
where the (linear) operator $\mathcal{L}$ is defined by
\begin{equation}
\label{eq:L}
    \mathcal{L}=\begin{pmatrix}
        -V_I-2\Phi_0^{R}\Phi_0^{I}& \qquad 
        \mu-V_R-\frac{1}{2}\partial_{xx}-\left[\Phi_0^{R}\right]^2
        -3\left[\Phi_0^{I}\right]^2
        \\[1.0ex]
        -\mu+V_R+\frac{1}{2}\partial_{xx}+3\left[\Phi_0^{R}\right]^2
        +\left[\Phi_0^{I}\right]^2&\qquad -V_I(x)+2\Phi_0^{R}\Phi_0^{I}
    \end{pmatrix},
\end{equation}
and where we have decomposed the wavefunction into its real and imaginary 
components according to $\Phi_0=\Phi_0^{R}+i\,\Phi_0^{I}$.
We now proceed to compute the spectrum ($\lambda$) of this linear operator
by using Fourier differentiation matrix approximations for the spatial 
derivative operators (cf.~Ref.~\cite{trefethen_spectral_2000}), 
in conjunction with the generalized Schur decomposition implemented 
in Matlab's \texttt{eig} function.
It is worth pointing out that the linear operator $\mathcal{L}$ in Eq.~\eqref{eq:L} is unitarily equivalent to the operator $\mathcal{M}$ arising from the linearization about the bright mode of the phase-transformed model, Eq.~\eqref{Inhomog-advective}. Specifically, the two operators are related by
\begin{equation}
\label{Unitary-equivalent}
\mathcal{M}=
\begin{pmatrix}
\cos\theta(x) & -\sin\theta(x) \\[0.5ex]
\sin\theta(x) & \cos\theta(x)
\end{pmatrix}^{-1}
\mathcal{L}
\begin{pmatrix}
\cos\theta(x) & -\sin\theta(x) \\[0.5ex]
\sin\theta(x) & \cos\theta(x)
\end{pmatrix},
\end{equation}
where the Wadati phase function is $\theta(x)=\int_{-\infty}^{x}\mathcal{W}(x')\, dx'$. Furthermore, the associated eigenmodes of $\mathcal{M}$ satisfy the relationship ${\bf v}=U^{-1}{\bf d} $, where $U$ is the rotation matrix and ${\bf d}$ are the eigenmodes of $\mathcal{L}$. Thus, $\mathcal{L}$ and $\mathcal{M}$ possess identical eigenspectra. 
This unitary equivalence justifies the indirect comparison between the PDE stability results obtained from $\mathcal{L}$ and the NCVA stability predictions developed for the phase-transformed model (see Sec.~\ref{NCVA-formulation}).
Concurrently to the PDE calculations, we also compute the fixed points $\vec{p}_0$ of 
the NCVA ODE system Eq.~\eqref{NCVA-ODE-system} (see Sec.~\ref{NCVA-formulation}) 
using standard fixed-point iterations. 
To assess the linear stability of these NCVA fixed points, we then evaluate the 
eigenvalues and eigenvectors of the corresponding $6\!\times\!6$ Jacobian matrix, 
$(d\vec{F}/d\vec{p})|_{\vec{p}_0}$.
Naturally, this can only aspire to describe a small subset of point spectrum modes
of the full PDE system.

\begin{figure}
    \centering
    \includegraphics[width=0.7\linewidth]{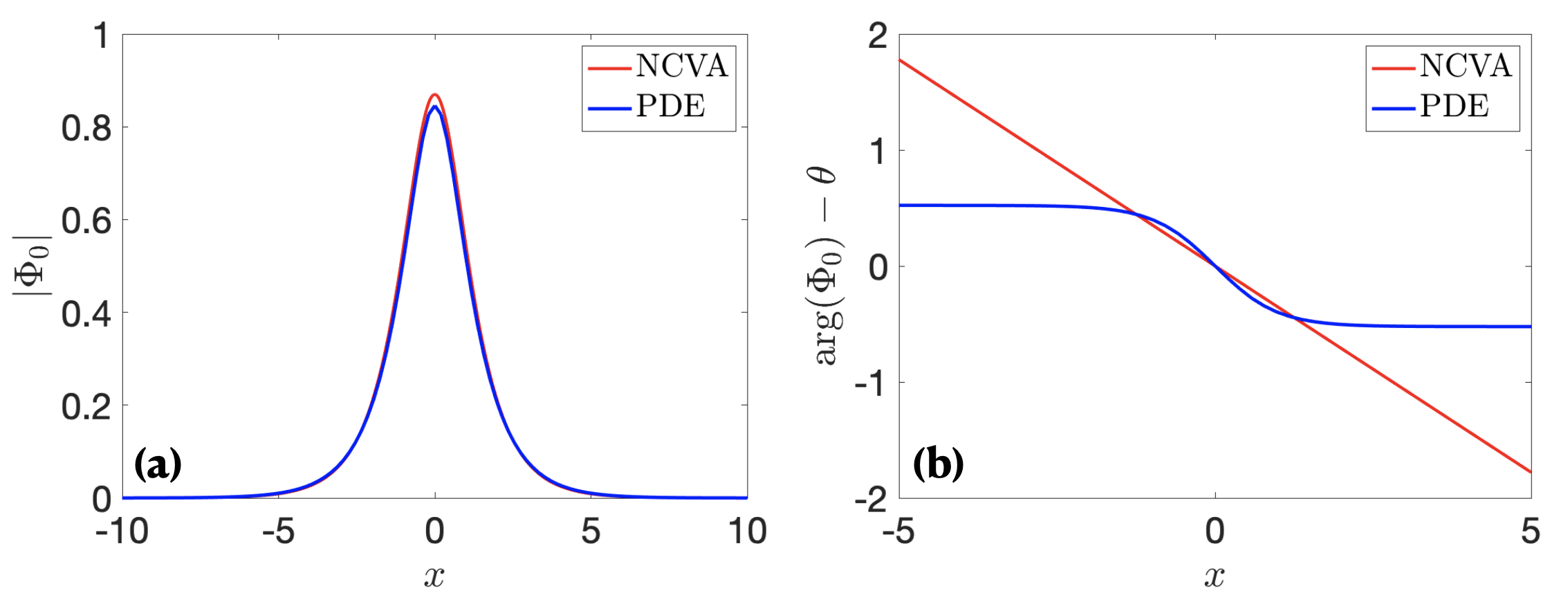}
    \caption{
    Representative bright mode for the NLS with a $\mathcal{PT}$-symmetric Wadati 
    potential (blue) compared with its NCVA counterpart (red) for $\alpha = 1$ 
    and $\mu = 0.5$. Panels (a) and (b) depict, respectively, the corresponding 
    wavefunction amplitudes and phases
    %
    }
    \label{fig:1}
\end{figure}

Figure~\ref{fig:1} depicts a representative bright soliton, for $\alpha=1$ and $\mu=0.5$, obtained from the PDE together with its corresponding low-dimensional NCVA approximation. The amplitude profiles, shown in panel (a), exhibit good agreement, with only minor discrepancies near the peak. The NCVA phase profile, which approximates the quantity $\argOperator(\Phi_0)-\theta$, preserves the $\mathcal{PT}$ symmetry of the PDE solution, with vanishing chirp $b=0$, and closely matches the PDE phase within the soliton core, as shown in panel (b). Noticeable differences arise only in spatial regions where the amplitude is negligible. The agreement between the PDE and NCVA amplitude profiles persists over most of the continuation interval in $\alpha$. The linear limit predicted by the NCVA occurs at $\alpha\approx1.88$, compared to $\alpha\approx1.74$ for the PDE bright-soliton branch.

\begin{figure}
    \centering
\includegraphics[width=0.4\linewidth]{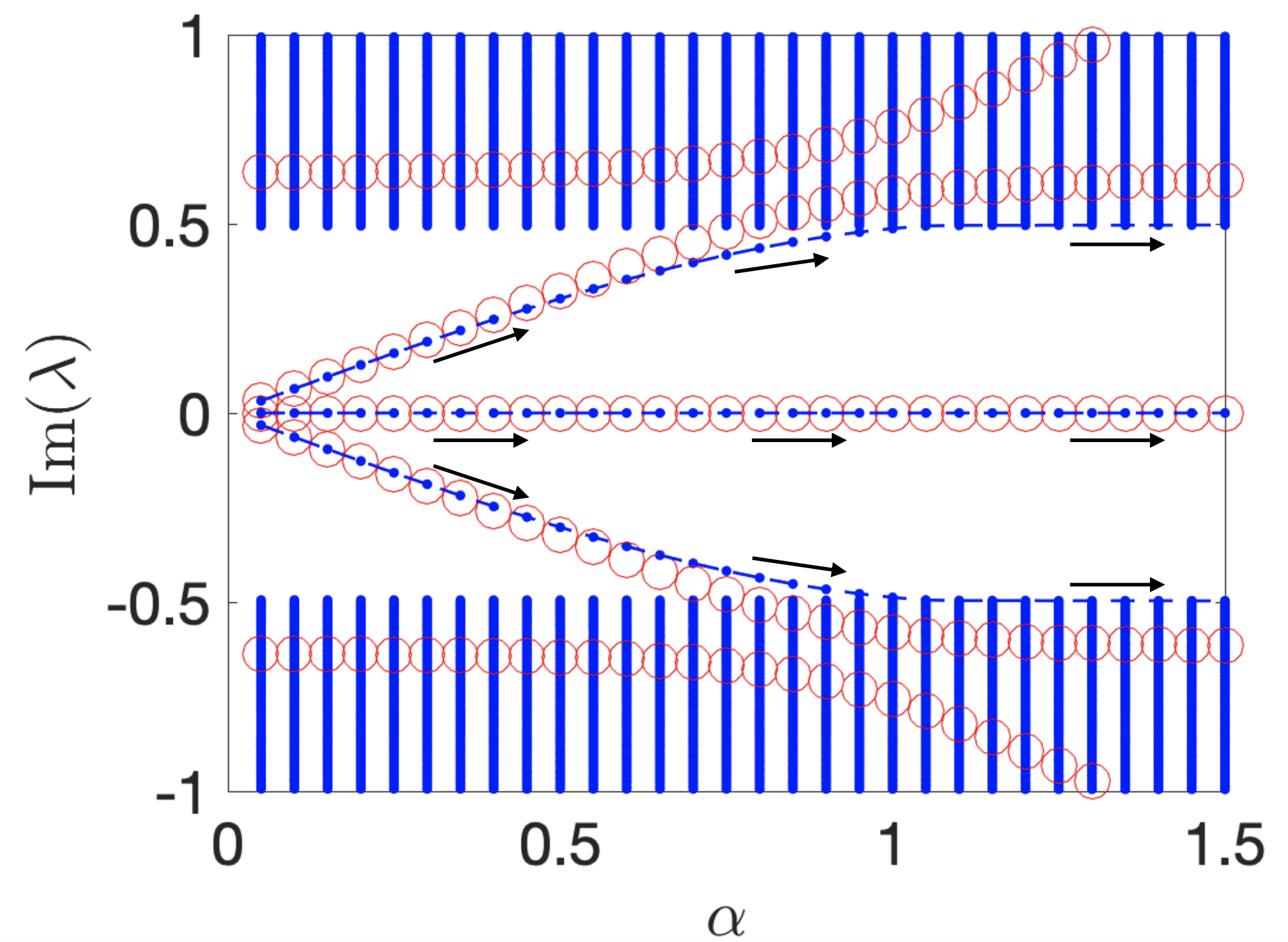}
\caption{Dependence of the purely imaginary eigenvalues, $\Imaginary(\lambda)$, on the 
strength parameter $\alpha$ of the $\mathcal{PT}$-symmetric Wadati potential. PDE eigenvalues are shown as blue dots and their NCVA counterparts as red circles. The translational eigenvalues, with arrows indicating their motion as $\alpha$ increases, asymptotically approach the edges of the continuous spectrum.}
    \label{fig:2}
\end{figure}

Next, we assess the fidelity of the NCVA in capturing the PDE eigenspectrum as a function of the potential  strength
parameter $\alpha$; see Fig.~\ref{fig:2}. A numerical computation of the eigenspectrum $\lambda$ of $\mathcal{L}$ (or, equivalently, $\mathcal{M}$) indicates linear stability, with $\Real(\lambda)\equiv 0$ over the entire existence interval in $\alpha$. Accordingly, Fig.~\ref{fig:2} shows the imaginary part of the eigenspectrum of $\mathcal{L}$ as a function of $\alpha$ (blue dots), displayed up to $\alpha=1.5$ for clarity, although the stability computations were performed over the entire existence interval of the PDE branch, up to its linear limit at $\alpha\approx1.74$. At $\alpha=0$, corresponding to the cubic NLS, four eigenvalues lie at the origin: one pair associated with translational invariance, and another {one associated} with phase (gauge) invariance. As $\alpha$ increases, the translational pair splits symmetrically from the origin, given the breaking of the symmetry due to the presence of the potential, and asymptotically approaches the edges of the continuous spectrum at $|\lambda|=\mu$. Thus, throughout the entire existence interval in $\alpha$, both the \textit{direct} linear stability analysis and the low-dimensional NCVA predict linearly stable nonlinear bright modes. Interestingly, the NCVA description also yields an additional pair of eigenvalues (given the six-dimensional nature
of the relevant representation) whose behavior appears to reflect, in a finite-dimensional sense, the influence of the PDE continuous spectrum. A caveat is warranted here: the low-dimensional approximation cannot genuinely capture the continuous spectrum or its associated delocalized eigenfunctions (cf.~Sec.~\ref{NCVA-formulation}). Nevertheless, it is noteworthy that this NCVA eigenvalue pair remains embedded within the continuous spectral band of the full PDE throughout the range of $\alpha$ considered, while progressively deviating from the NCVA approximation of the translational eigenvalue pair. This behavior (in conjunction with that of the pair 
showcasing the breaking of translational invariance) bears a suggestive resemblance to the asymptotic approach of the PDE translational eigenvalues toward the edges of the continuous spectrum.

It is relevant to mention that accurately computing the spectrum of a non-self-adjoint matrix differential operator on the entire real line is notoriously difficult, frequently resulting in clusters of spurious complex eigenvalues due to numerical artifacts~\cite{guba2011continuous,sandstede2001stability,deconinck_computing_2006,deconinck2006transverse}. {The strategy we employ to mitigate these spurious effects is to enlarge the computational domain and increase the resolution by employing more Fourier modes ---in particular, this led us to observe the \textit{algebraic} convergence to the true continuous spectrum. Specifically, for the computations reported here, we use the domain $[-100,100]$ with $N=2048$ Fourier modes.}

\begin{figure}
    \centering
    \includegraphics[width=0.75\linewidth]{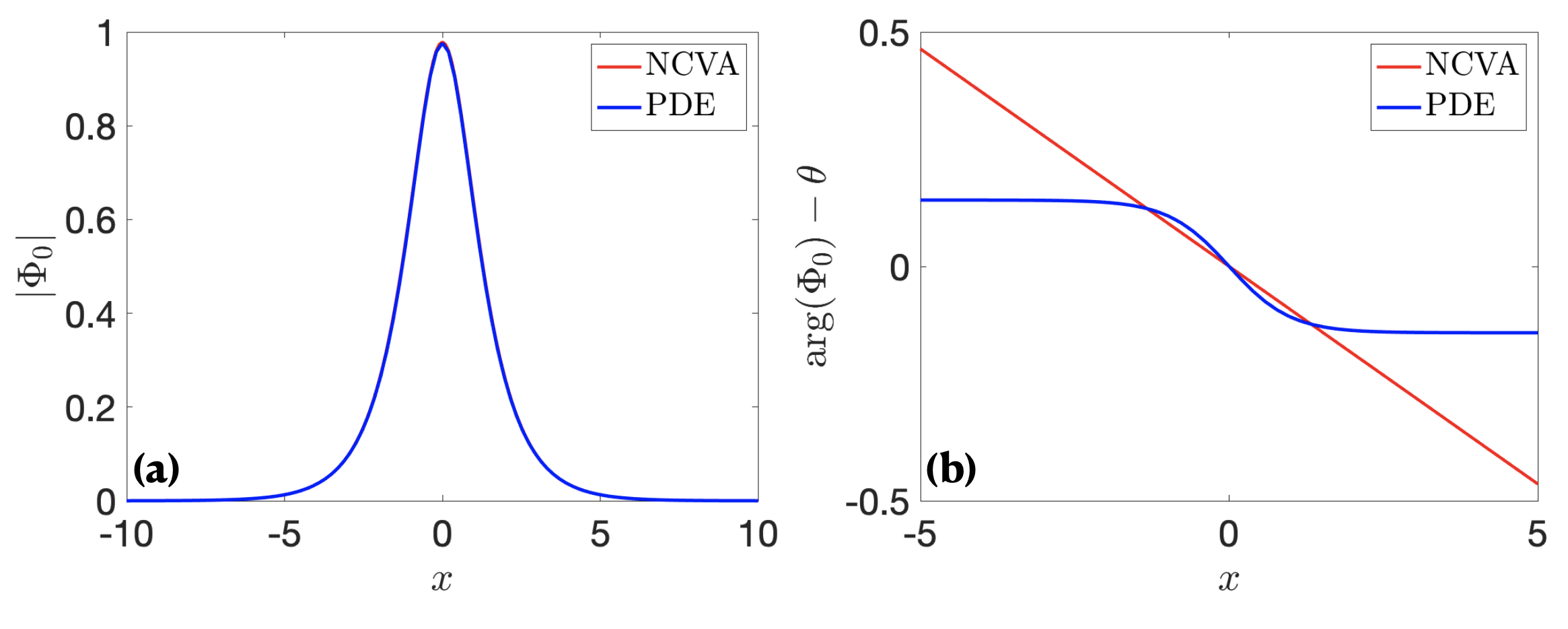}
    \caption{
    Same as in Fig.~\ref{fig:1} for a representative bright soliton for a non-Wadati, $\mathcal{PT}$-symmetric potential. 
    }
    \label{fig:PT-symmetry}
\end{figure}

\subsection{Non-Wadati $\mathcal{PT}$-symmetric potentials}
\label{Non-Wadati-PT}

We now study a representative family of non-Hermitian optical systems 
possessing $\mathcal{PT}$ symmetry, but lacking the Wadati structure. 
To this end, we consider the one-parameter family of potentials defined in Eq.~\eqref{PT-symmetry-but-no-Wadati}. A key difference from Sec.~\ref{Wadati-PT-case} is that, in the present case, the parameters defining the $\mathcal{PT}$ symmetry need not lie within the so-called ``unbroken $\mathcal{PT}$-symmetric phase.'' In particular, the linear eigenvalue problem
$$
\left(\frac{1}{2}\frac{d^2}{dx^2}+V(x)\right)\varphi=\Lambda\varphi
$$
may possess complex eigenvalues when the gain-loss component of the potential is sufficiently strong relative to its conservative counterpart (cf.~Ref.~\cite{ahmed2001real} for a discussion of the related Scarf-II potential). At the outset, we remark that the $\mathcal{PT}$-symmetry-breaking threshold of the underlying linear problem is numerically estimated to be $\alpha_{\rm PT}\approx 0.62$. %
%
As Fig.~\ref{fig:1-1}(b) evidences, the family of $\mathcal{PT}$-symmetric soliton solutions exists for all values of the propagation constant $\mu$ larger than the value corresponding to the linear limit, where the norm of the solution tends to zero.

With this in mind, we examine the stability properties of the nonlinear bright solitons for the representative propagation constant $\mu=0.5$. Notably, the $\mathcal{PT}$-symmetry-breaking threshold of the underlying linear problem need not coincide with the onset of instability of the nonlinear bright-soliton branch.
%

\begin{figure}[htbp]
    \centering
    \includegraphics[width=0.8\linewidth]{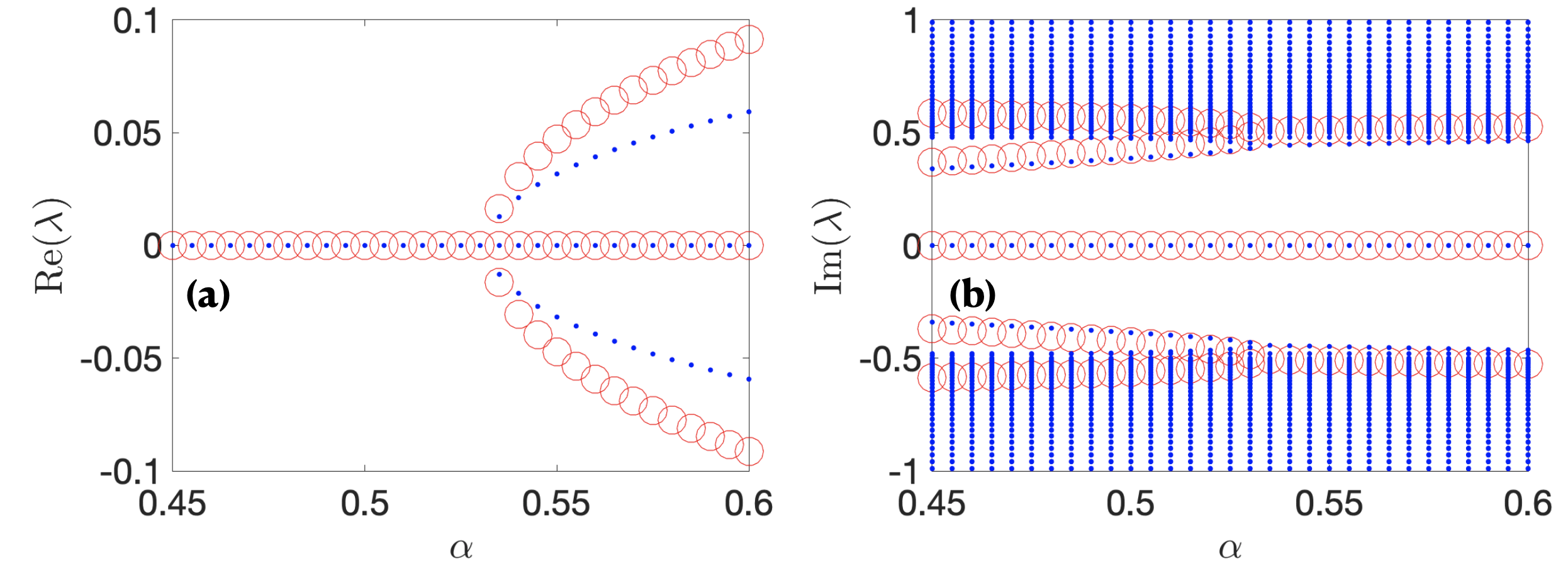}
    \caption{Dependence of the spectrum [real part in panel (a) and imaginary part in 
    panel (b)] on the depth parameter $\alpha$ of the $\mathcal{PT}$-symmetric potential. PDE results are shown as blue dots and NCVA eigenvalues as red circles. For smaller $\alpha$, the spectrum is purely imaginary, with translational eigenmodes approaching the edges of the continuous spectrum. 
    %
    }
    \label{fig:Spectra-PT-symmetry}
\end{figure}

We compute $\mathcal{PT}$-symmetric bright solitons for $0<\alpha\leq 0.6$ using the numerical continuation scheme discussed in Sec.~\ref{Wadati-PT-case}. Here, we restrict the continuation to a narrow interval beyond the instability threshold of the nonlinear bright solitons. Concurrent to the PDE solutions, we also study the fixed points (and stability) of the NCVA 
system~\eqref{NCVA-ODE-system}.
In Fig.~\ref{fig:PT-symmetry}, we depict a  representative case ($\alpha=0.55$), 
lying in the unbroken $\mathcal{PT}$-symmetric phase where the PDE and NCVA wavefunctions 
show good agreement in amplitude and reasonable agreement in phase [i.e., the phase of 
the soliton governed by Eq.~\eqref{phi-time}], particularly in the region of the soliton core. 

We now examine the linear stability spectrum $\lambda(\alpha)$ predicted by the PDE and assess the extent to which it is captured by the NCVA description. The resulting spectral properties as a function of the potential depth parameter $\alpha$ are shown in Fig.~\ref{fig:Spectra-PT-symmetry}, with the real and imaginary parts of the eigenvalues displayed in panels (a) and (b), respectively. As previously discussed in Sec.~\ref{Wadati-PT-case}, the finite-dimensional NCVA possesses only discrete eigenvalues; nevertheless, in the context of the Hopf bifurcation, one of these eigenvalues effectively tracks the relevant edge of the PDE continuous spectrum. For values of $\alpha<0.535$, the spectrum is entirely imaginary, with the translational eigenmodes approaching the edges of the continuous spectrum as $\alpha$ increases. At $\alpha\approx 0.455$, an additional discrete mode detaches from the edge of the continuous spectrum and subsequently approaches the translational eigenmodes with increasing $\alpha$; see panel (b). At $\alpha\approx0.535$, these modes collide, giving rise to a complex quartet of eigenvalues through a Hopf-type bifurcation~\cite{chernyavsky2018krein}, thereby signaling the onset of oscillatory instability of the 
bright soliton. Remarkably, the NCVA accurately predicts the location of this bifurcation to within the second decimal place, besides respecting the quartet symmetry beyond. Beyond the bifurcation point, however, a small but progressively growing discrepancy between the NCVA and PDE growth 
rates, associated with the real part of the relevant eigenvalue, emerges as $\alpha$ increases.

\begin{figure}[htbp]
    \centering
    \includegraphics[width=\linewidth]{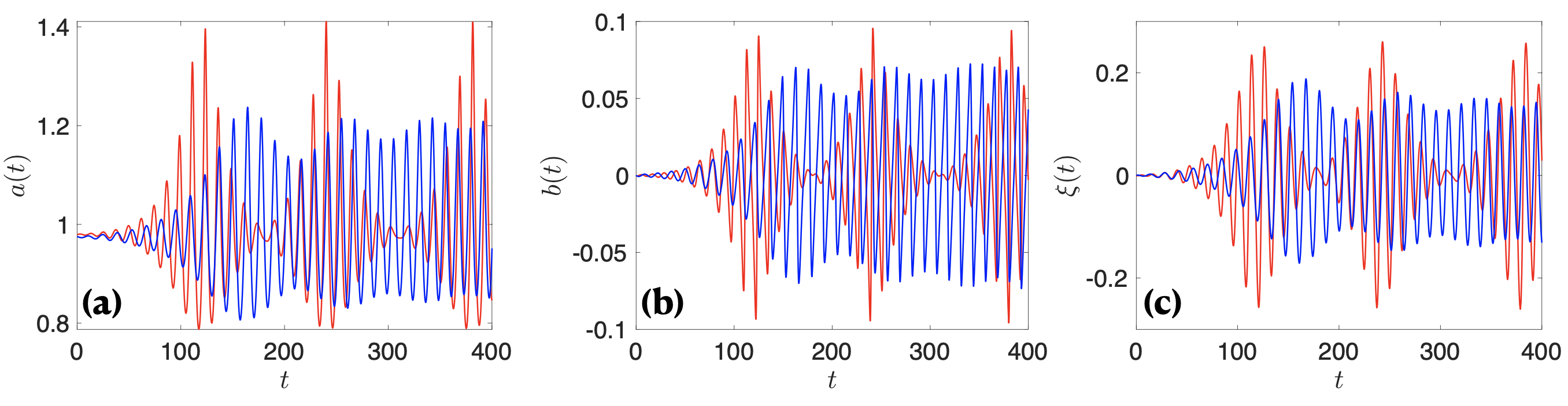}
    \caption{Destabilization dynamics of the bright soliton in the $\mathcal{PT}$-symmetric 
    but non-Wadati case for $\alpha= 0.55$. 
    Shown are the temporal evolution of (a) the amplitude $a(t)$, (b) the chirp $b(t)$, 
    and (c) the center-of-mass position $\xi(t)$ of the perturbed bright soliton.%
    The PDE dynamics, projected onto the solitonic manifold, are shown in blue, 
    while the corresponding NCVA dynamics are shown as solid red lines. 
    }
    \label{fig:ncva-pde-PT-noWadati}
\end{figure}

In Fig.~\ref{fig:ncva-pde-PT-noWadati} we depict a typical evolution of perturbed, unstable bright solitons in the PDE [Eq.~\eqref{phi-time}] and NCVA [Eqs.~\eqref{NCVA-ODE-system}] within a $\mathcal{PT}$-symmetric potential of the non-Wadati type, for $\alpha=0.55$. The full PDE dynamics, governed by Eq.~\eqref{phi-time}, is projected onto the solitonic manifold and shown in blue, while the corresponding NCVA predictions are depicted as solid red lines. At early times, the trajectories of both models closely follow the growth rates predicted by their respective linear stability analyses. The close agreement between the projected PDE and NCVA dynamics in this regime suggests that the tangent space of the solitonic manifold is sufficiently rich to capture the dominant unstable PDE eigenmode. As time advances, robust quasi-periodic dynamics emerge within the NCVA framework, which initially appear to qualitatively mimic the projected PDE dynamics. Closer inspection, however, reveals an important discrepancy: the PDE solution emits 
radiation, a feature that cannot be represented by the spatially localized NCVA ansatz. Consequently, while the NCVA predicts persistent quasi-periodic dynamics confined to the solitonic manifold (see Appendix~\ref{PT-noWadati-Appendix}), the PDE dynamics are accompanied by weak cw radiation, reflecting degrees of freedom beyond those retained by the finite-dimensional variational approximation. 
Over long times (tracked up to $t=1000$; not shown here), the projected PDE evolution exhibits qualitatively different behavior, with the amplitude and inverse-width parameters undergoing oscillations about a secularly increasing, phase-locked trend.
On the other hand, the oscillations in the chirp intensify over longer times.

\begin{figure}[htbp]
    \centering
    \includegraphics[width=0.75\linewidth]{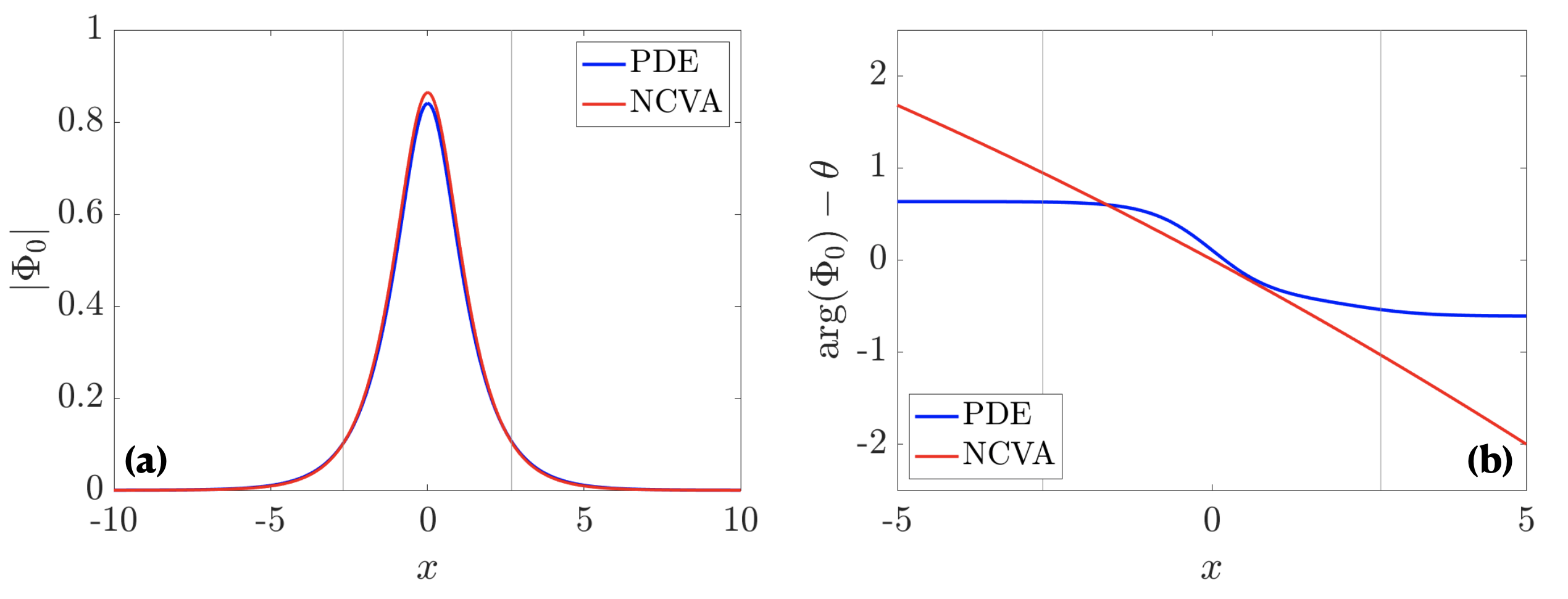}
    \caption{
    Comparison of a representative bright mode in a non-$\mathcal{PT}$-symmetric 
    Wadati potential with its NCVA counterpart, using parameters $\alpha = 1$ 
    and $\mu = 0.5$ (red: NCVA; blue: PDE):
    (a) The wavefunction amplitude demonstrates good agreement with the six-dimensional NCVA description.
    (b) The shifted phase of the wavefunction exhibits a substantially more pronounced, 
    and even qualitative, disagreement with the PDE solution than in the case where
    $\mathcal{PT}$ symmetry is also present. 
    This discrepancy persists even within the spatial region where the soliton 
    amplitude is appreciable (see thin vertical lines depicting the width of the
    bright soliton defined by $|\Phi_0|>1/e^2$). 
    %
    }
    \label{fig:wadati-ptasymmetry}
\end{figure}

\subsection{Non-$\mathcal{PT}$-symmetric, Wadati potentials}
\label{Non-PT-Wadati}

We now consider the case that exhibits Wadati structure without imposing 
$\mathcal{PT}$ symmetry. In particular, we focus on real, single-humped 
Wadati potential functions $\mathcal{W}(x)$, i.e., functions that vary 
monotonically up to a single extremum and monotonically in the opposite 
sense thereafter. {Importantly, single-humped potentials within this class lack an even parity about their extremum.} Through the correspondence between the associated 
Zakharov--Shabat and Schrödinger spectral problems~\cite{KlausShaw2002}, such single-humped 
functions are known to generate Wadati potentials whose discrete eigenvalues 
are purely real~\cite{tsoy2014stable}, in contrast to the $\mathcal{PT}$-only 
case (Sec.~\ref{Non-Wadati-PT}).

We note ---from the outset--- the existence of continuous families of solitons 
as parametrized by the soliton frequency ($\mu$), as shown in Fig.~\ref{fig:1-1}(c).
{For concreteness, we fix $\Gamma=0.1$ and $x_0=2.75$ in
Eq.~\eqref{WADATI-no-PT}. Starting from the cubic NLS bright soliton with
propagation constant $\mu=0.5$, we  perform numerical continuation in
the parameter $\alpha$ over the interval $0<\alpha\lesssim 1.7$ to compute
the bright-soliton solutions of Eq.~\eqref{nonlinear-eigenmodes}.} The existence of continuous families of bright solitons in non-$\mathcal{PT}$-symmetric Wadati potentials is a remarkable consequence of an additional conserved quantity associated with the stationary-mode equations~\cite{konotop2014families,NixonYang2016}. This conserved quantity effectively reduces the number of independent constraints governing smooth bright solitons, thereby closing the stationary problem for a prescribed $\mu$ and potential ($\alpha$, $\Gamma$) parameters.

The upper end of this continuation branch corresponds to the relevant linear limit. 
As expected, the resulting non-centered bright solitons are not $\mathcal{PT}$-symmetric.
Now, we also seek to assess whether the six-dimensional NCVA provides an appropriate
minimal model. To this end, we concurrently compute the fixed points of
Eq.~\eqref{NCVA-ODE-system} and examine the agreement between the resulting
solitonic-manifold approximation and the corresponding PDE solution. As in Sec.~\ref{Wadati-PT-case}, the NCVA slightly overpredicts the soliton
amplitude, which, in turn, leads to an approximate $9\%$ extension of the
predicted existence interval in $\alpha$. A representative comparison for
$\alpha=1$ is shown in Fig.~\ref{fig:wadati-ptasymmetry}. 
The most notable discrepancies, both qualitative and quantitative, arise in the
asymmetric phase profile of the PDE soliton; and thus cannot be adequately represented by the low-dimensional quadratic phase ansatz; see panel (b).
Moreover, for the computations presented in this section, the NCVA bright-soliton 
fixed points satisfy $b_0<0$ and $\xi_0>0$, in contrast to the 
$\mathcal{PT}$-symmetric cases, for which $\xi_0=b_0=0$. 
\begin{figure}
[htbp]
    \centering
    \includegraphics[width=0.7\linewidth]{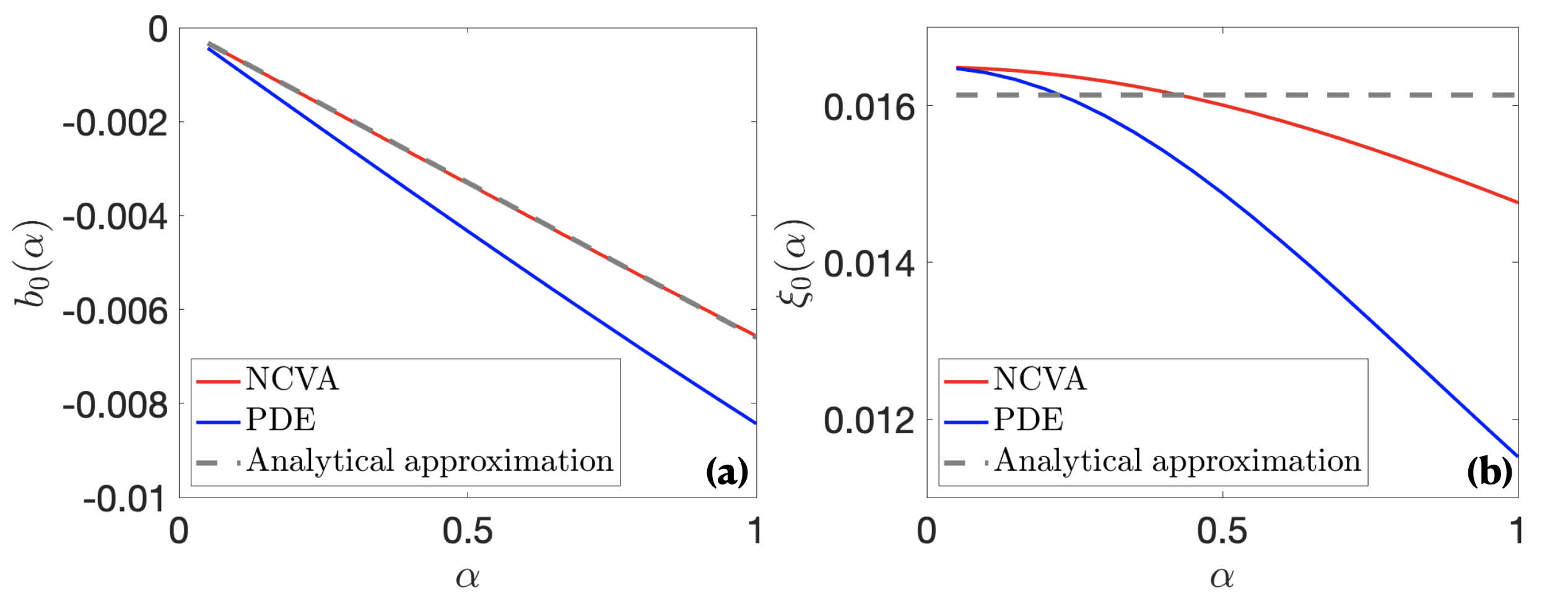}
    \caption{The NCVA parameters (red curves) are compared with their approximate analytical expressions (gray dashed lines) for the non-$\mathcal{PT}$-symmetric Wadati potentials in Eq.~\eqref{WADATI-no-PT}, obtained under the assumption of an approximately constant soliton width. The corresponding PDE parameters are also overlaid (blue curves), obtained by projecting the PDE solutions onto the NCVA ansatz through a best-fit procedure. For the parameter set $x_0=2.75$, $\Gamma=0.1$, and $\mu=0.5$, the soliton width remains nearly constant throughout the entire existence interval of the NCVA bright-soliton family. Shown here is the variation with $\alpha$ of (a) the chirp parameter and (b) the soliton center.}
    \label{soliton-chirp-center-1}
\end{figure}
We probe the NCVA bright soliton parameters analytically, assuming that $w_0(\alpha)\approx w_0(0)$, i.e., that the soliton width remains approximately constant as $\alpha$ varies. This approximation is supported by the computed solutions over a broad range of $x_0$ corresponding to single-humped potentials, for which the variation of $w_0$ remains weak. This behavior can be understood from the fact that the dominant localization scale is set primarily by the underlying potential. Furthermore, for the discussion below, we focus on the 
{case with} $\mu=0.5$, for which $w_0(0)=1$. We then examine how the weak asymmetry, characterized by $\Gamma\ll1$ in Eq.~\eqref{WADATI-no-PT}, induces a nonzero chirp, $b_0\sim\Gamma b_1$, and a small displacement of the soliton center, $\xi_0\sim\Gamma\xi_1$. At $\mathcal{O}(1)$ in the asymmetry parameter, the fixed-point condition $I_{\chi}=0$ is satisfied identically by parity considerations, since $\int_{-\infty}^{\infty}\sech^4(x)\tanh(x) dx=0$. At $\mathcal{O}(\Gamma)$, it yields the expression for the soliton center
\begin{equation}
\label{center-of-mass}
\xi_1=
\frac{
\displaystyle\int_{-\infty}^{\infty}
\sech^2(x-x_0)\sech^2(x)\tanh(x)\,dx
}{
\displaystyle\int_{-\infty}^{\infty}
\sech^4(x)\left[1-3\tanh^2(x)\right]\,dx
}
=
\frac{30\sech^2(x_0)}{8\tanh^4(x_0)}\left[{x_0(3-\tanh^2(x_0))-3\tanh(x_0)}\right],
\end{equation}
whose exact evaluation is standard, by leveraging the 
identity $\tanh(x-x_0)=[\tanh(x)-\tanh(x_0)]/[1-\tanh(x)\tanh(x_0)]$.  
In the limit of weak asymmetry, $x_0\ll1$, the expansion
$\xi_1(x_0)\approx x_0-(10/21)x_0^3+\cdots$
reveals an approximately linear dependence of the soliton-center correction on $x_0$, and reduces to the result $\xi_1(0)=0$ in the $\mathcal{PT}$-symmetric limit.
In particular, the formula in Eq.~\eqref{center-of-mass} indicates that 
the soliton center is shifted slightly right (since we are using $x_0>0$). 
This prediction agrees reasonably well with the bright-soliton center for $\alpha\leq1$, with deviations remaining within approximately $10\%$. At the same time, the center itself is found numerically to depend only weakly on $\alpha$; see Fig.~\ref{soliton-chirp-center-1}(b), where the NCVA result is shown by the red solid line and its analytical approximation by the gray dashed line. Across the existence interval, the NCVA inverse-width parameter deviates from unity by as much as approximately $15\%$, which likely accounts for the increasing discrepancy of the analytical prediction for the center at larger $\alpha$. Nevertheless, the weak overall dependence of $\xi_0$ on $\alpha$ can be understood directly from the mass-balance condition. 
In particular, owing to the weak dependence of $w_0$ on $\alpha$, the condition $I_{\chi}(\xi_0,w_0)=0$ itself depends only weakly on $\alpha$, \textit{through} its dependence on the inverse-width parameter. This, in turn, explains the weak dependence of the soliton center $\xi_0$ on $\alpha$.

It is also useful to highlight the increasing discrepancy, with increasing $\alpha$, between the NCVA and PDE soliton centers. The latter is estimated by performing a least-squares best fit of the NCVA solitonic ansatz to the PDE solution, with the global phase parameter fixed to $\chi_0=0$, consistently with the phase convention adopted for the NCVA fixed point; see blue curve in Fig.~\ref{soliton-chirp-center-1}(b). On another note, the fixed-point relation $b_0\approx-(3/(2a^2\pi^2))I_b$ yields, at $\mathcal{O}(\Gamma)$,
\begin{equation}
\label{Chirp-correction}
b_1=
-\frac{3\alpha}{\pi^2}
\left[
\int_{-\infty}^{\infty}
\sech^2(x-x_0)G(x)\,dx
-\frac{2\pi^2}{45}\xi_1
\right],
\end{equation}
where $G(x)=x^2\sech^2(x)\tanh(x)$, and whose evaluation involves a convergent Mellin-type integral~\cite{gradshteyn1988tables}. Carrying out the integration, we obtain the approximate expression for the chirp,
$\Gamma b_1=-\frac{8\Gamma\alpha x_0^2}{15\pi^2}\xi_1<0$. This prediction is nearly coincident with the exact NCVA soliton chirp; see Fig.~\ref{soliton-chirp-center-1}(a). In contrast, as in the case of the soliton center, the chirp extracted from the PDE solution (blue line) exhibits an increasing discrepancy from the NCVA prediction as $\alpha$ is increased.

\begin{figure}[htbp]
    \centering
    \includegraphics[width=0.4\linewidth]{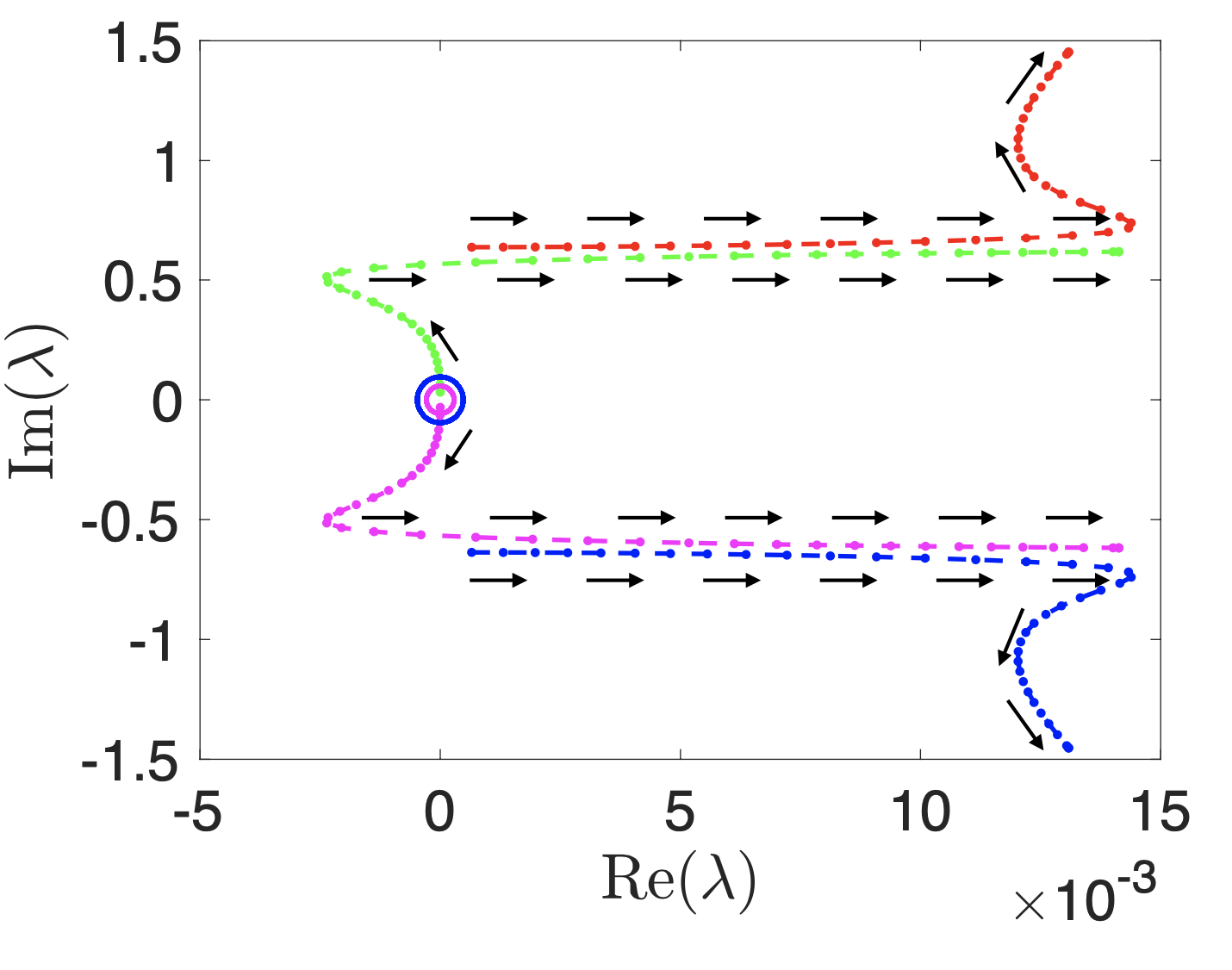}
    \caption{
    The trajectory of the NCVA eigenvalues $\lambda$ as the potential strength $\alpha$
    is increased (see arrows), for the no-$\mathcal{PT}$-symmetric, Wadati potential case.
    The eigenvalues associated with the phase invariance stay fixed to the origin 
    (blue and magenta circles), while the pair associated with the translational 
    invariance of the $\alpha=0$ case depart from the origin (green and magenta dots).  
    This pair eventually predicts a maximum instability growth rate 
    [$\Real(\lambda)\sim 10^{-2}$] corresponding to the zero-amplitude limit.   
    }
    \label{fig:placeholder2}
\end{figure}

Next, as concerns the PDE stability spectrum, we find that
$i\lambda_{\alpha}\in\mathbb{R}$ ({besides spurious finite-domain numerical 
artifacts, progressively disappearing as the numerical domain gets larger}) 
throughout the entire interval of
$0<\alpha\lesssim 1.7$ considered, indicating that the bright solitons remain spectrally
stable. The resulting stability spectra is therefore qualitatively similar
to that observed in Sec.~\ref{Wadati-PT-case}, and therefore not separately shown. 
We next examine,  see Fig.~\ref{fig:placeholder2}, the eigenvalues of the 
$6\!\times\!6$ NCVA Jacobian as a function of the potential strength 
$0<\alpha\lessapprox 1.86$, noting at the outset that the continuous 
spectrum cannot be represented within this finite-dimensional framework. 
The arrows in the figure indicate the trajectories of the eigenvalues in
the complex-$\lambda$ plane as $\alpha$ is increased. In addition, the instability 
is exacerbated as the zero-amplitude limit is approached. Finally, the NCVA 
spectrum lacks the Hamiltonian-like quartet symmetry observed in the full 
linearization spectra of non-$\mathcal{PT}$-symmetric Wadati 
systems~\cite{yang2016stability}. We examine this issue analytically 
in the Appendix~\ref{NCVA-instability-Appendix}, wherein it becomes apparent that 
the spurious instability is due to the limitations inherent to the 
chosen variational ansatz.

\subsection{Non-Wadati and non-$\mathcal{PT}$ symmetric potentials}
\label{No-wadati-no-PT}

In this section, we investigate non-Hermitian potentials lacking both $\mathcal{PT}$ symmetry and Wadati structure. In particular, we consider the family of potentials defined in Eq.~\eqref{VR_VI-non-conservative}, where $\beta\ll1$ parametrizes the departure from the Wadati limit. Throughout this section, we additionally restrict the asymmetry parameter $\Gamma$ to be small, such that $\mathcal{W}(x)=\sqrt{2V_R(x)}$ is a single humped potential function.
The existence and properties of bright solitons in dissipative settings (besides the $\mathcal{PT}$ and Wadati potentials) have attracted considerable interest, particularly in connection with the persistence, or breakdown, of continuous families of stationary solutions parametrized by 
$\mu$~\cite{zezyulin2022nonexistence,kominis2019continuous}. 
Such investigations generally indicate that, for generic non-Hermitian potentials, bright solitons occur only as isolated solutions, since, for a given, prescribed $\mu$, the stationary-mode problem is effectively overdetermined. {As in the PDE bright-soliton problem, the NCVA fixed-point system was shown in Sec.~\ref{NCVA-formulation} to be generically overdetermined.}

\begin{figure}[htbp]
    \centering
    \includegraphics[width=0.75\linewidth]{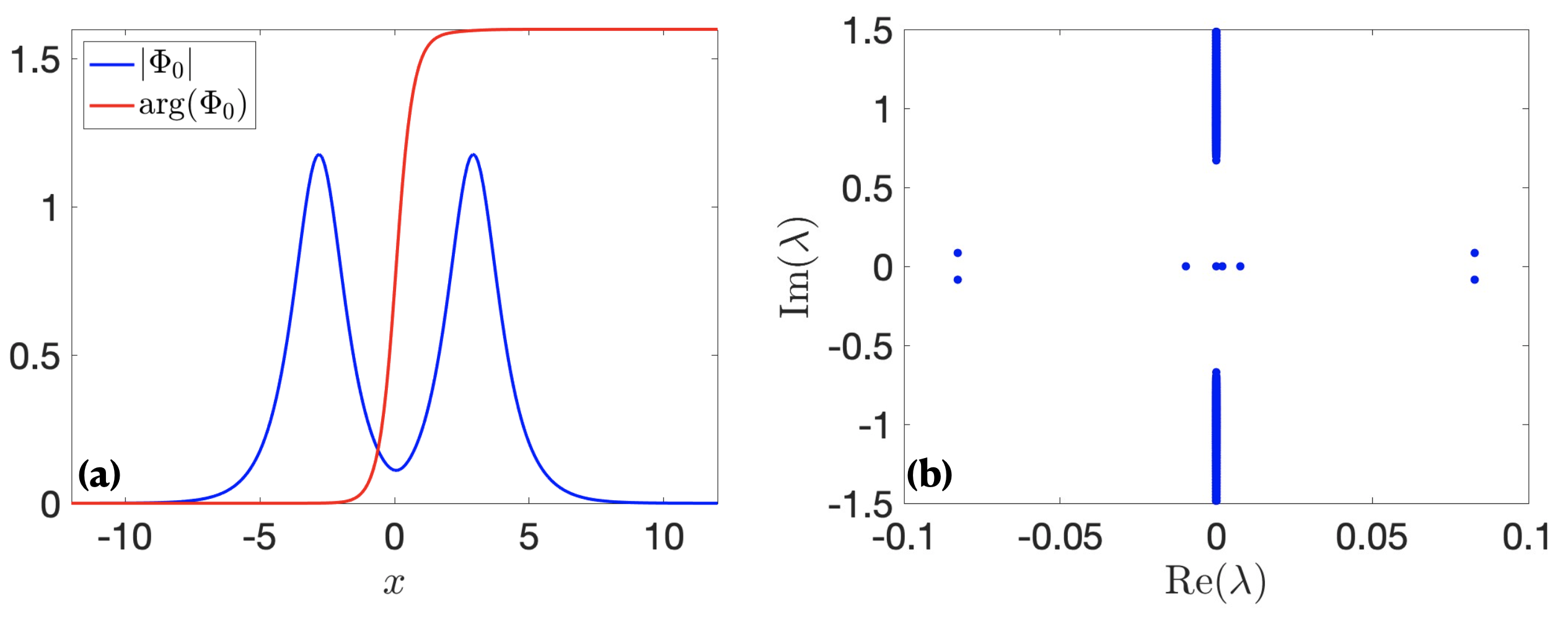}
    \caption{
    Generic bright solitons supported by non-$\mathcal{PT}$-symmetric and non-Wadati 
    potentials [see Eq.~\eqref{VR_VI-non-conservative}] with parameters $\alpha=0.1$, 
    $\Gamma =0.1$ and $\beta=0.05$. The associated propagation constant is found to 
    be $\mu\approx0.694$. Panel (a) depicts the amplitude (blue) and phase (red) 
    profiles of the bright soliton while panel (b) depicts its corresponding 
    linear stability spectrum evidencing its unstable nature. 
    Unlike the Wadati case, the spectrum does not display the quartet symmetry.}
    \label{fig:mode}
\end{figure}

To compute bright solitons supported by generic non-Hermitian waveguides, 
we employ a standard shooting method; see, e.g., Ref.~\cite{konotop2014families}. 
First, we express the stationary wavefunction in polar form, $\Phi_0(x)=A(x)\exp[i\Theta(x)]$, 
{where $A(x)$ 
is the wave amplitude and $u(x)=\Theta'(x)$ is the local hydrodynamic velocity (phase gradient). This}
yields the following 
system of coupled ODEs 
{of} hydrodynamic form,
\begin{subequations}
\label{Hydrodynamic-equations-bright-soliton}
\begin{eqnarray}
A''(x)&=&2\mu A(x)+A(x)u^2(x)-2V_R(x)A(x)-2A^3(x),
\label{Hydrodynamic-equations-bright-soliton_A}
\\[1.0ex]
\label{Hydrodynamic-equations-bright-soliton_B}
u'(x)&=&-2u(x)\frac{A'(x)}{A(x)}-2V_I(x).
\end{eqnarray}
\end{subequations}
%
Given the decay of the potential in the far field, the system can be solved exactly
there, yielding $A(x)\rightarrow r_{\pm}\exp(\mp\sqrt{2\mu}\,x)$ as $x\rightarrow\pm\infty$. The two far-field amplitudes $r_{\pm}$, together with the propagation constant $\mu$, constitute three unknown parameters, which are determined by requiring the solutions integrated from the two asymptotic boundaries to match smoothly at $x=0$. This gives the three matching conditions
$$
A(0^+)=A(0^-),\qquad
A'(0^+)=A'(0^-),\qquad
u(0^+)=u(0^-).
$$

Having determined the unknown parameters $r_\pm$ and $\mu$, the ODE system in 
Eq.~\eqref{Hydrodynamic-equations-bright-soliton} can be integrated forwards, 
after which the corresponding wavefunction is reconstructed as 
$\Phi_0(x)=A(x)\exp[i\int_{-\infty}^{x}u(x')dx']$. 
We performed an extensive search for stationary solutions in the weakly 
non-Wadati regime, $\beta\ll1$, but were unable to identify single-humped bright 
solitons, i.e., solutions whose amplitude possesses a single local maximum. 
Instead, double-humped stationary modes were found generically. Moreover, the 
separation between the two amplitude maxima was observed to increase relative 
to the $\beta=0$ Wadati limit. A representative example of this double-humped 
localized solution and its stability spectrum are shown in Fig.~\ref{fig:mode}.
We note {that}, as it is the case in the representative case presented 
in Fig.~\ref{fig:mode},  bright solitons in the non-$\mathcal{PT}$ and 
non-Wadati case are generically unstable. 
Furthermore, unlike the $\mathcal{PT}$-symmetric or Wadati case(s), 
the quartet symmetry in the spectrum is not preserved; see panel (b) in
Fig.~\ref{fig:mode}. 

We next investigate the existence of bright solitons to the 
NCVA with non-conservative terms (where $\mu$ is unknown) in Eq.~\eqref{phi-time} 
with $\mathcal{R}=-i\mathcal{W}\phi_x-i(\beta/2)\mathcal{W}_x\phi$. In this vein, 
after suitable manipulation of the $I_{\chi}=0$ equation (i.e., an integration by parts) we obtain 
$$2(1-\beta)a_0^2 w_0 \int_{-\infty}^{\infty} \mathcal{W} \sech^2(w_0(x-\xi_0))\tanh(w_0(x-\xi_0)) dx=0.$$ 
This yields the usual relationship $\xi_0\sim \Gamma \xi_1$ as was derived in Eq.~\eqref{center-of-mass}, for a small asymmetry parameter $\Gamma$. Now we examine the integral 
\begin{equation}
\label{Ixi-sectord}
I_{\xi}=2\beta a_0^2 b_0 \int_{-\infty}^{\infty} \mathcal{W}'(x) (x-\xi_0)\sech^2(w_0(x-\xi_0)) dx=0.
\end{equation}
The equation necessitates that either $b_0=0$ or the integral vanishes.
Note that if $\mathcal{W}(x)$ has an unique maximum at $x=x_m$ (as a single-humped function), 
and utilizing  $I_{\chi}=0$, we can show the integral is negative, since 
\[
\int_{-\infty}^{\infty} \mathcal{W}'(x)(x-\xi_0)
\sech^2\left(w_0(x-\xi_0)\right)\,dx
+\frac{\xi_0-x_m}{(1-\beta)a_0^2}I_{\chi}
=
\int_{-\infty}^{\infty} \mathcal{W}'(x)(x-x_m)
\sech^2\left(w_0(x-\xi_0)\right)\,dx
<0.
\]
This suggests that $b_0=0$. However, the condition $F_w=0$ gives $b_0=-(3w_0^3/(2a_0^2\pi^2))I_b$, where
\begin{align}
I_b=2a_0^2\Bigg[
&w_0(1-\beta)\int_{-\infty}^{\infty}\mathcal{W}(x)(x-\xi_0)^2
\sech^2\!\left[w_0(x-\xi_0)\right]
\tanh\!\left[w_0(x-\xi_0)\right]\,dx 
\nonumber
\\
\nonumber
&+\beta\int_{-\infty}^{\infty}\mathcal{W}(x)(x-\xi_0)
\sech^2\!\left[w_0(x-\xi_0)\right]\,dx
\Bigg].
\end{align}
Assuming $\beta\sim\Gamma\ll1$, the leading-order contribution to $I_b$ is inherited from the $\beta=0$ Wadati limit and is generically nonzero, while the departure from the Wadati
type of potential introduces only $\mathcal{O}(\beta)$ corrections to $I_b$. Since $b_0\sim\alpha\Gamma b_1$ [see discussion of Eq.~\eqref{Chirp-correction}] at $\beta=0$, continuity for sufficiently small $\beta$ therefore suggests that $I_b$ remains nonzero, and hence that $b_0\neq0$. 
This is incompatible with the condition $b_0=0$ inferred from $F_c=0$
{[which can be deduced from $I_{\xi}=0$, see Eq.~\eqref{Ixi-sectord}]},
indicating a possible incompatibility between the $F_c=0$ and $F_w=0$ fixed-point equations and, consequently, the possible absence of NCVA fixed points in the regime $\beta\sim\Gamma\ll1$.

\begin{figure}[htbp]
    \centering
    \includegraphics[width=0.7\linewidth]{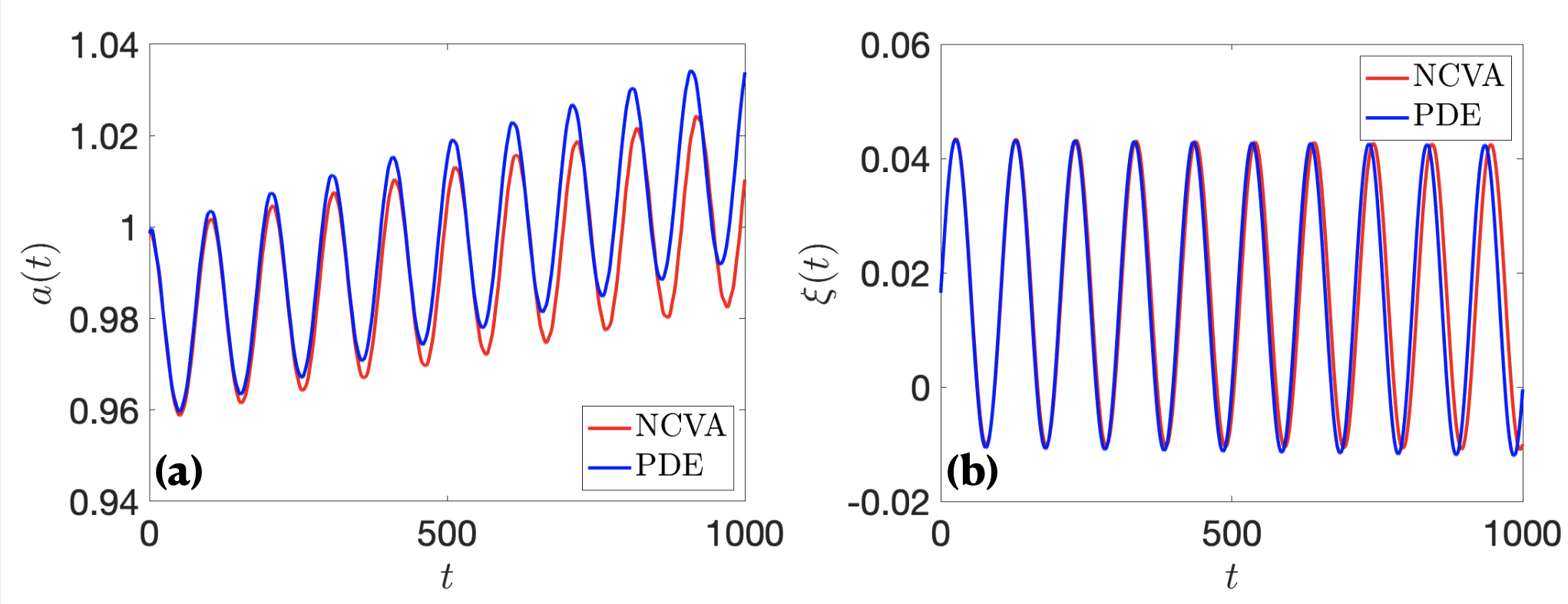}
    \caption{The dynamics of Wadati bright solitons when the Wadati potential is weakly perturbed so as to no longer possess the Wadati property. The parameters characterizing the underlying Wadati potential are $\alpha=0.1$, $\Gamma=0.1$, $x_0=2.75$, and propagation constant $\mu=0.5$, while the perturbation parameter is $\beta=0.05$. The PDE dynamics (blue curves) are projected onto the NCVA ansatz through a best-fit procedure and compared with the corresponding NCVA dynamics (red curves). (a) The peak amplitude of the bright soliton oscillates in phase with the inverse-width parameter, with the envelope of both oscillations exhibiting approximately linear growth. (b) The soliton center executes approximately periodic oscillations about the unperturbed equilibrium position $\xi_0$ corresponding to $\beta=0$.}
    \label{fig:NL_DYNAMICS_WADATI}
\end{figure}

Another intriguing aspect corresponds to the fate of the (single-humped) Wadati bright 
solitons ($\beta=0$) when the Wadati structure of the underlying potential is perturbed. 
This of course has implications for their robustness, and thus, experimental observability. 
We generally observe that Wadati bright solitons ($\beta=0$) appear to play an important 
role in the dynamics for $\beta\ll1$ as the mode executes small amplitude oscillations 
about its unperturbed center, while the envelope of oscillations is also increasing 
slightly in {the projected soliton} amplitude; see Fig.~\ref{fig:NL_DYNAMICS_WADATI} 
for a representative example (for $\alpha=0.1$, $\Gamma=0.1$, and $\beta=0.05$). 
Similar behavior has also been reported for ``pseudomodes" of the 
so-called W-dW potentials~\cite{zezyulin2022nonexistence}. 
Notably, the projected soliton amplitude and width evolve nearly in phase and remain closely locked throughout the dynamics, while the oscillations of the projected soliton center are approximately $\pi/2$ out of phase with them (see below).

Interestingly, when the potential is perturbed away from the Wadati limit ($\beta\neq0$, Fig.~\ref{fig:NL_DYNAMICS_WADATI}), the dynamics of the NCVA solitons initialized from the Wadati limit exhibit {very good} agreement with the PDE dynamics. {Both the (projected) PDE and NCVA width parameters (not shown) oscillate in phase with the amplitude parameter and possess nearly the same temporal profile (i.e., $a(t)\approx w(t)$), consistent with the relationship $d/dt\left(\int_{\mathbb{R}} |\overline{\phi}|^2 dx\right)=d/dt(2a(t)^2/w(t))\approx 2\dot a(t)$. Moreover, the {NCVA} soliton center oscillates about its unperturbed ($\beta=0$) equilibrium position, with a phase difference of nearly $\pi/2$ radians relative to the soliton amplitude, akin to the projected PDE dynamics in Fig.~\ref{fig:NL_DYNAMICS_WADATI}(b).}

\section{Conclusions and scope for future work}
\label{Conclusions}

In this work, we developed a unified classification of the non-Hermitian potentials
based on their connection to the $\mathcal{PT}$ and the Wadati property. 
This in turn yielded four distinct settings: systems possessing both properties, systems with $\mathcal{PT}$ symmetry alone, systems with Wadati structure alone, and systems possessing neither. 
We examined the existence and stability of bright solitons in non-Hermitian optical systems. 
Within the above classification, we also assessed the extent to which a low six-dimensional variational reduction can faithfully capture the existence and spectral stability of the corresponding soliton families within each category.

When at least one of these structural properties is present, continuous families 
of bright solitons parametrized by their (temporal) frequency $\mu$ are found 
to exist, a feature that is also captured by the variational approximation. 
At the NCVA level, this can be understood in part from the fact that the corresponding algebraic system governing the fixed points is exactly determined. 
%
For the $\mathcal{PT}$-symmetric families examined here, the NCVA reproduces the relevant PDE stability behavior with good accuracy.
For the case in which Wadati structure is present in the absence of 
$\mathcal{PT}$ symmetry, the NCVA successfully reproduces the bright-soliton intensity, but fails to accurately capture the asymmetric features of phase structure. A spurious instability in the NCVA reduced model is also witnessed, perhaps due to the inadequacy of the variational ansatz itself (see Appendix~\ref{NCVA-instability-Appendix}). It would be interesting to develop an improved solitonic ansatz that could incorporate a potential-adapted phase correction (while keeping the same number of parameters). Of interest would also be to check if an NCVA with such an improved ansatz could also preserve the quartet symmetry of the linearized stability spectrum as entailed by the Wadati structure. 

For the near-Wadati family considered here, our numerical searches yield isolated, typically double-humped and unstable localized states, and we have not identified corresponding single-humped stationary modes in the parameter regime examined. This observation is consistent with the overdetermined system of equations governing the existence of NCVA fixed points. Moreover, weak departures from the Wadati limit lead, within the NCVA framework, to an incompatibility that (generically) precludes single-humped bright-soliton fixed points. Whether this obstruction at the variational level can be translated into a more concrete statement regarding the existence of single-humped NLS solitons under weak departures from Wadati structure remains an interesting question for future work. Yet another extension would be to generalize the NCVA ansatz so as to accommodate multi-humped localized states.

Further interesting directions include investigating how the NCVA framework may be extended to accommodate symmetry-broken branches in $\mathcal{PT}$-symmetric Wadati potentials~\cite{yang2014symmetry}. Other natural avenues for future work include developing variational approximations for heteroclinic and homoclinic states of the defocusing NLS with Wadati potentials~\cite{chandramouli2026homoclinic}, and also carrying out a more comprehensive study of non-Hermitian scattering phenomena in both focusing and defocusing settings not only
in one, but also in higher dimensions where associated studies are far more sparse.

\section{Appendix}
\appendix

\subsection{Non-conservative variational approximation}
\label{NCVA-Appendix}

The non-conservative variational approximation (NCVA) represents a semi-analytical 
method~\cite{rossi2020non} to dynamically reduce the NLS equation ---in the presence
of gain and loss terms--- to an effective ``particle'' picture described by a set of 
coupled ODEs on the soliton parameters. To employ the NCVA, we consider two sets of 
coordinates $\phi_1$ and $\phi_2$, that are fixed at an initial 
time ($t_i$), but are not fixed at the final time ($t_f$). Upon application of
variational calculus for a non-conservative system, both paths are set equal, 
$\phi_1= \phi_2$, and identified with the physical path $\phi$, i.e., ``the physical 
limit'' (PL). The action functional for $\phi_1$ and $\phi_2$ is defined as the 
total line integral of the difference of the Lagrangians ($\mathcal{L}$) between the paths 
with the sum of the line integral of the functional ${\cal R}$ which describes the 
the non-conservative, path-dependent forces: 
\begin{eqnarray}
S &=& \int_{t_i}^{t_f}  \left(\mathcal{L} (\phi_1, \phi_{1,t}, \phi_{1,x}, \ldots,t) 
- \mathcal{L}(\phi_2, \phi_{2,t}, \phi_{2,x}, \ldots,t) + \mathcal{R} \right)\, dt,
\label{Action}
\end{eqnarray}
Here, the $t$ and $x$ subscripts represent the partial derivatives
with respect to the corresponding variables.
The action in Eq.~\eqref{Action} defines a non-conservative Lagrangian density 
$\mathcal{\tilde{L}} \equiv \mathcal{L}_1 - \mathcal{L}_2 + \mathcal{R}$.
Then, the equations of motion at the physical limit yield
$$
    {\frac{\partial}{\partial t}\left(\frac{\delta \mathcal{L}}{\delta {\phi_t^{*}}}\right)=\frac{\delta \mathcal{L}}{\delta {\phi^{*}}}+\left[\frac{\delta \mathcal{R}}{\delta \phi_{-}^{*}}\right]_{\rm PL}},
$$
where $\delta$ denotes the Fréchet derivatives and $\phi_{-}\equiv\phi_1-\phi_2$. 
We thus obtain the Euler-Lagrange (EL) equation for the conservative 
terms and all non-conservative terms are lumped into 
$[{\delta \mathcal{R}}/{\delta \phi_{-}^{*}}]_{\rm PL}$. 
Notably, this field-theoretic formulation of the non-conservative problem 
has not deployed any approximation yet. 
Now, we proceed to solve the above EL equations using a suitable ansatz to 
approximately describe solitons in the non-conservative Lagrangian formulation.
An effective method to study the dynamics of the solution is to restricted it on a 
solitonic manifold, i.e., assume an ansatz $\phi\equiv \overline{\phi}(x,\vec{p}(t))$
with time-dependent parameters $\vec{p}(t))$, 
for which the modified EL equations for the averaged Lagrangian 
${\overline{L}=\int_{-\infty}^\infty \overline{\mathcal{L}}\, dx}$ 
result in the ODE system
$$
\frac{d}{dt}\left( \frac{\partial \overline{L}}{\partial \dot{\vec{p}}} \right)-\frac{\partial \overline{L}}{\partial \vec{p}} -  I_{\vec{p}} = 0,
$$
where the integrals $I_{\vec{p}}$\, are given by 
$I_{\vec{p}}\equiv \int_{-\infty}^\infty \left[{\partial \mathcal{\overline{R}}}/{\partial \vec{p}_- }\right]_{\rm PL} d x$ 
and the overline indicates evaluation on the solitonic ansatz.

\subsection{Periodic Orbits and Bifurcations in the NCVA Dynamics of the $\mathcal{PT}$-Symmetric, non-Wadati case}
\label{PT-noWadati-Appendix}

We comment further on the long-time dynamics of the finite-dimensional NCVA description observed in Fig.~\ref{fig:ncva-pde-PT-noWadati}, particularly the emergence of multiscale, quasi-periodic breather dynamics. To elucidate their origin, we compute the branch of periodic breather solutions bifurcating from the Hopf bifurcation of the stationary bright soliton identified in Fig.~\ref{fig:Spectra-PT-symmetry}, and subsequently examine the stability and secondary bifurcations of these periodic orbits. At $\alpha=0.55$, the stationary NCVA solution lies beyond the Hopf bifurcation point. The resulting complex quartet $\{\lambda,\lambda^*,-\lambda,-\lambda^*\}$ contains the eigenvalue $\lambda\approx0.0472+0.5110i$. We denote the corresponding complex eigenvector by $\vec{V}=\vec{a}+i\vec{b}$, where $\vec{a},\vec{b}\in\mathbb{R}^6$. Periodic orbits of the NCVA system~\eqref{NCVA-ODE-system} are then computed as fixed points of the time-$T$ map,
$ \chi^{T}[\vec{\mathcal{P}}(0)]=\vec{\mathcal{P}}(0)$,
where $\chi^{T}[\cdot]$ denotes the flow map of the dynamical system over a time interval $T$
and $\vec{\mathcal{P}}(0)$ denotes the initial point on the periodic orbit. An additional
\textit{phase-pinning condition}, $a(0)=a_0$, is appended to this shooting problem, 
thereby selecting the \textit{a priori} unknown breather propagation constant $\mu$. 
We perform continuation in the period $T$, with the initial period informed by the 
imaginary part of the Hopf eigenvalue, $T_0={2\pi}/\Imaginary(\lambda)$. The shooting 
computation is initialized using 
$\vec{\mathcal{P}}_{\rm breather}(0) \approx \vec{p}_0+\epsilon\vec{a}$,
where $\epsilon\ll1$ and $\vec{p}_0$ denotes the fixed point of the NCVA ODE system. 
To determine the stability of the resulting periodic orbits, we compute their 
Floquet multipliers $\Lambda$, i.e., the eigenvalues of the monodromy matrix 
obtained by linearizing the time-$T$ map about each periodic orbit.
 
Stability is determined by the location of these multipliers relative to the unit circle in the complex plane. Importantly, the periodic breather branch is itself found to lose stability through a Neimark--Sacker bifurcation of the associated Poincar\'e map~\cite{kuznetsov1998elements}, at $T\approx12.37$. 
%
%
This secondary instability of the periodic breather thus provides a natural mechanism (and a qualitative explanation) for the observed beating and multiscale NCVA quasi-periodic dynamics in Fig.~\ref{fig:ncva-pde-PT-noWadati}. Thus, the long-time NCVA dynamics should 
 be interpreted as dynamics associated with the instability of the breather branch beyond its Neimark--Sacker bifurcation.

\subsection{NCVA solitons in non-$\mathcal{PT}$-symmetric, Wadati potentials}
\label{NCVA-instability-Appendix}

\begin{figure}[htbp]
    \centering
    \includegraphics[width=0.35\linewidth]{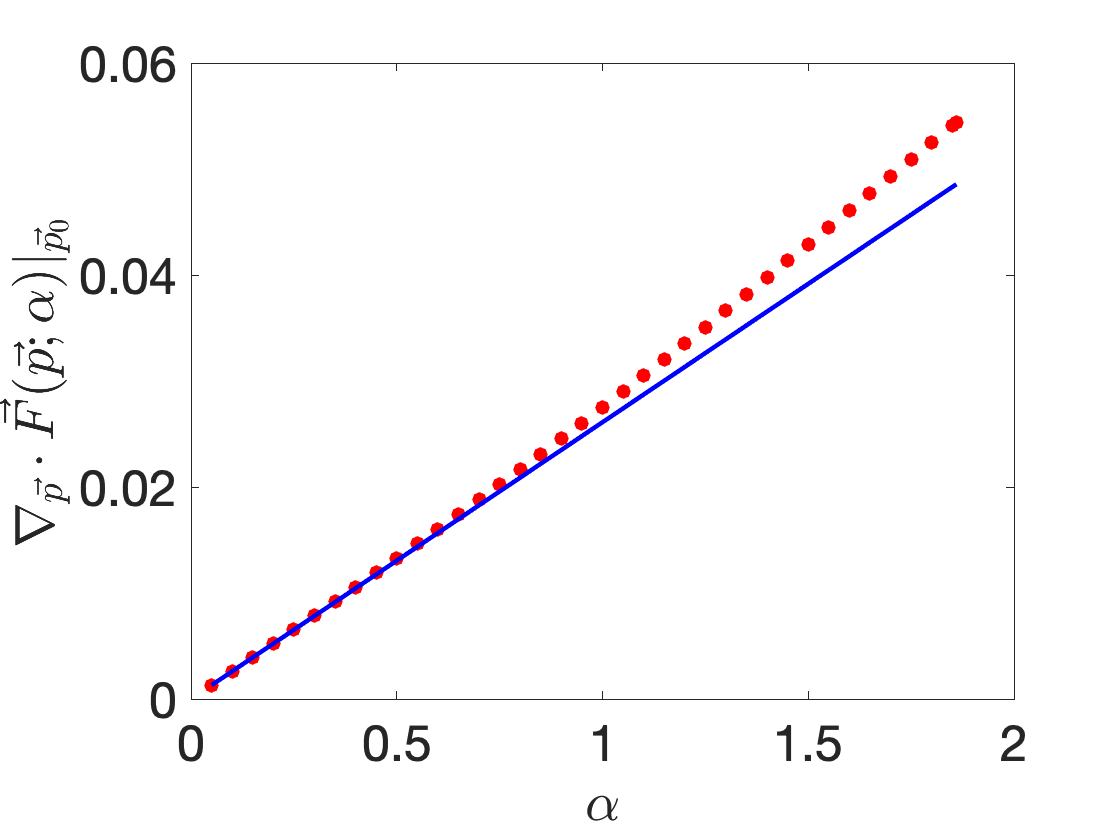}
    \caption{The trace of the NCVA Jacobian for the non-$\mathcal{PT}$-symmetric Wadati potential (red dots: numerical evaluation of Jacobian trace, blue line: analytical prediction) shown for the entire existence regime $0<\alpha\lesssim 1.86$ for $\mu=0.5$.} 
    \label{fig:wadati_ncva_failure}
\end{figure}

\begin{figure}[htbp]
    \centering
    \includegraphics[width=0.65\linewidth]{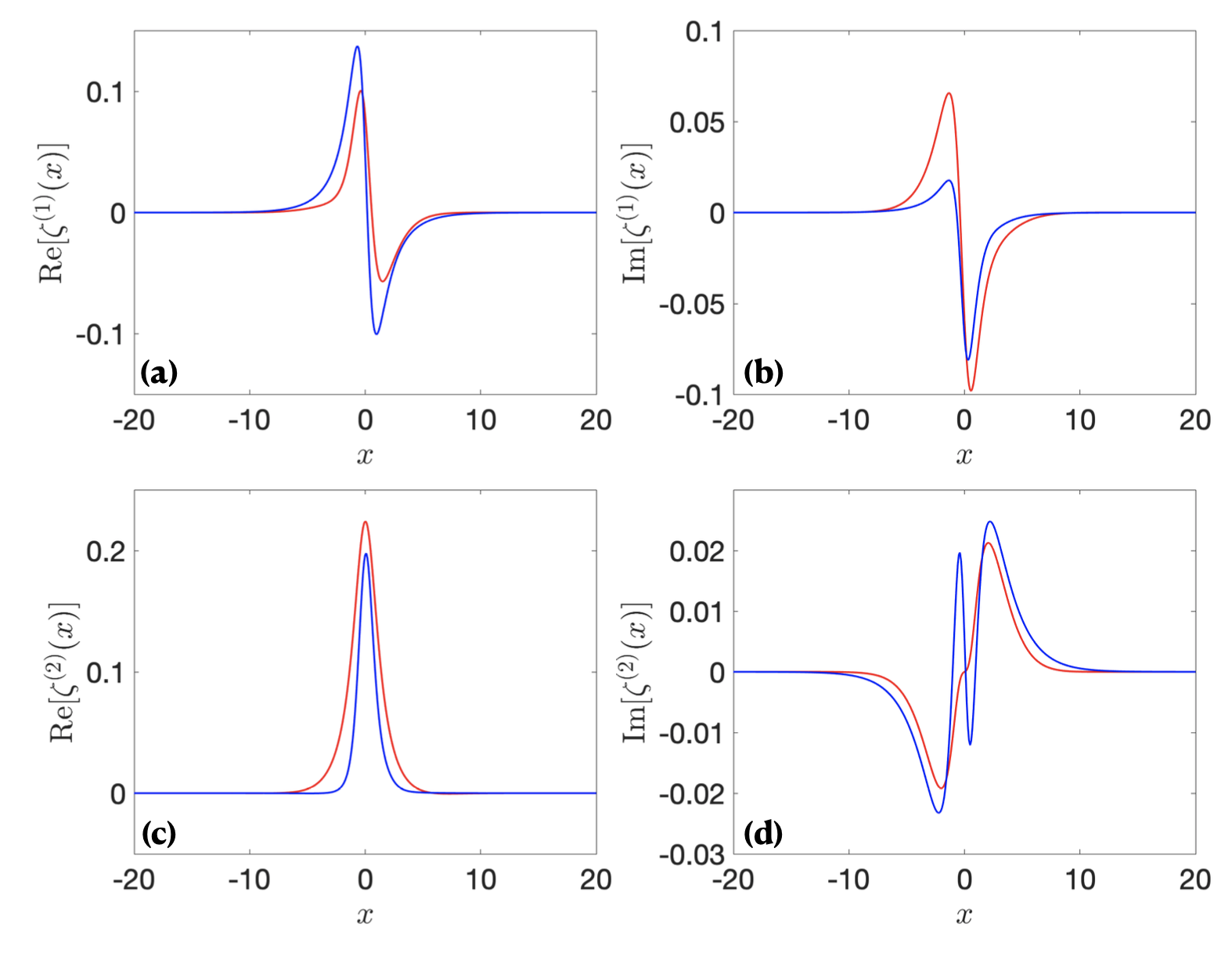}
    \caption{
    The components of the eigenmode {$[\zeta^{(1)}(x),\zeta^{(2)}(x)]^T$} associated with 
    the eigenvalue $\lambda\approx 0.35 i$ of the linear stability operator $\mathcal{M}$ 
    corresponding to Eq.~\eqref{Inhomog-advective}. The PDE eigenmode (blue) is compared 
    with its NCVA counterpart (red).
    }
    \label{fig:EIGENFUNCTIONS}
\end{figure}

In this section, we investigate the origin of the spurious NCVA instability identified in Sec.~\ref{Non-PT-Wadati}, and discuss how it may arise from limitations inherent to the variational ansatz considered. To begin our discussion, we recall the identity
$\Trace(J)=\nabla_{\vec{p}}\cdot\vec{F}(\vec{p};\alpha)|_{\vec{p}_0}
=\sum_{i=1}^{6}\lambda_i$.
For a neutrally stable fixed point with a purely imaginary spectrum, the trace must therefore vanish, whereas asymptotic stability requires $\Trace(J)<0$. More generally, the Hamiltonian-like quartet symmetry $\{\lambda,-\lambda,\lambda^*,-\lambda^*\}$ likewise requires
$\Trace(J)=\nabla_{\vec{p}}\cdot\vec{F}(\vec{p};\alpha)|_{\vec{p}_0}=0$. 
To estimate this quantity analytically, we note at the outset, that gauge invariance implies
${\partial F_{\chi}}/{\partial \chi}|_{\vec{p}_0}=0$.
Furthermore, evaluation of the NCVA equations at the fixed point yields
${\partial F_a}/{\partial a}|_{\vec{p}_0}\equiv 0$.
Finally, for Wadati potentials, the fixed-point condition $F_c(\vec{p}_0)=0$, together with $I_{\xi}\equiv 0$ [cf.~the discussion below Eq.~\eqref{NCVA-ODE-system}], yields $\left.\partial F_c/\partial c\right|_{\vec{p}_0}\equiv 0$. Thus, the trace of the $6\!\times\!6$ Jacobian is given by
$$
\left.\nabla_{\vec{p}}\cdot\vec{F}(\vec{p};\alpha)\right|_{\vec{p}_0}
=
\left[
\frac{\partial F_b}{\partial b}
+
\frac{\partial F_{\xi}}{\partial \xi}
+
\frac{\partial F_w}{\partial w}
\right]_{\vec{p}_0}.
$$

As the localization scale is dominantly set by the complex potential (Sec.~\ref{Non-PT-Wadati}), we assume that $w_0(\alpha)\approx w_0(0)$, i.e., that the soliton width remains approximately constant as $\alpha$ varies. Furthermore, we focus on the parameter $\mu=0.5$, for which $w_0(0)=1$.

With the expressions for the chirp {$\Gamma b_1=-\frac{8\Gamma\alpha x_0^2}{15\pi^2}\xi_1<0$} and soliton center in Eq.~\eqref{center-of-mass}, we obtain the three contributions to the Jacobian trace, which, upon retaining terms up to $\mathcal{O}(\Gamma)$, yields
\begin{eqnarray}
\frac{\partial F_b}{\partial b}
&\sim&
\frac{6\alpha\Gamma}{\pi^2}
\left[
\int_{-\infty}^{\infty}
\sech^2(x-x_0)x\sech^2(x)\,dx
-\frac{2}{3}\xi_1
\right]
\nonumber
\\
&&=
\frac{6\alpha\Gamma}{\pi^2}\left[2x_0\csch^2(x_0)
\left(x_0\coth(x_0)-1\right)-\frac{2}{3}\xi_1\right],
\\[1.0ex]
\label{F-xi}
\frac{\partial F_{\xi}}{\partial \xi}
&\sim&
\alpha\Gamma
\left[
\int_{-\infty}^{\infty}
\frac{d}{dx}\left[\sech^2(x-x_0)\right]
x\sech^2(x)\tanh(x)\,dx
+\frac{4}{15}\xi_1
\right],
\\[1.0ex]
\label{F-w}
\frac{\partial F_w}{\partial w}
&\sim&
\alpha\Gamma
\left[
\int_{-\infty}^{\infty}
\sech^2(x-x_0)H(x)\,dx
+\frac{4}{5}\xi_1
\right],
\end{eqnarray}
where  
$$H(x)=
\frac{1}{2}x\sech^2(x)\left[1-3\tanh^2(x)\right]
-\frac{24}{\pi^2}x^2\sech^2(x)\tanh(x)
-\frac{6}{\pi^2}x^3\sech^2(x)\left[1-3\tanh^2(x)\right].$$ {The integral on the RHS of Eq.~\eqref{F-xi} can be evaluated exactly as $4x_0\csch^2(x_0)
\left[
x_0\coth(x_0)\left(1+3\csch^2(x_0)\right)
-2-3\csch^2(x_0)
\right]$}, while the integral on the RHS of Eq.~\eqref{F-w} yields

\begin{eqnarray}
\int_{-\infty}^{\infty}
\sech^2(x-x_0)H(x)\,dx
&=&
\frac{1}{\pi^2}\csch^2(x_0)
\Big\{
4x_0(3x_0^2+\pi^2)
\left[
x_0\coth(x_0)
\left(1+3\csch^2(x_0)\right)
-2-3\csch^2(x_0)
\right]
\nonumber
\\
\nonumber
&&+
(16x_0^2+2\pi^2)
\left[
3\coth(x_0)
-x_0\left(2+3\csch^2(x_0)\right)
\right]
\Big\}.
\end{eqnarray}

For $x_0=2.75$ and $\Gamma=0.1$, we compare the numerically extracted Jacobian 
trace (positive across the entire range in $\alpha$) with the constant-$w$ 
approximation and find agreement to within $11\%$. This level of agreement is 
reasonable given that $w(\alpha)$ itself varies by up to $16.7\%$ over the 
full range of $\alpha$ considered. 
Figure~\eqref{fig:wadati_ncva_failure}, depicting the trace of the NCVA Jacobian,
indicates that, in the vicinity of the fixed point $\vec{p}_0$, the reduced 
flow is locally volume-expanding on the solitonic manifold. 

Beyond the spectral discrepancies discussed above, we also examine the
differences between the NCVA and PDE representations of the 
eigenmodes associated with the eigenvalue pair that moves away from the
origin upon the breaking of translational invariance for $\alpha>0$.
A representative comparison at $\alpha=0.6$ is shown in
Fig.~\ref{fig:EIGENFUNCTIONS}. At this parameter value, the PDE eigenvalue
is $\lambda\approx0.35i$, while the corresponding NCVA eigenvalue is
$\lambda\approx-0.0011+0.3781i$. Thus, although the NCVA eigenvalue has
already acquired a small nonzero real part, it has not yet crossed into
the right-half plane. The NCVA approximation to the corresponding {two-component} 
eigenmode {$[\zeta^{(1)}(x),\zeta^{(2)}(x)]^T$ of the linear stability operator Eq.~\eqref{Unitary-equivalent}} is constructed as
\[
\left[
\sum_{i=1}^{6} v_i^{(j)}
\partial_{p_i}\Real\left(\phi(x,\vec{p}_0)\right),
\;
\sum_{i=1}^{6} v_i^{(j)}
\partial_{p_i}\Imaginary\left(\phi(x,\vec{p}_0)\right)
\right]^T,
\]
where $\vec{v}^{(j)}$ denotes the associated NCVA (normalized) eigendirection. As shown in Fig.~\ref{fig:EIGENFUNCTIONS}, appreciable quantitative discrepancies are present in all four components, while panel~(d) additionally exhibits a clear qualitative mismatch between the NCVA and PDE eigenmodes. In particular, this indicates that the tangent space spanned by the NCVA manifold may be insufficient to accurately represent the corresponding PDE eigenmodes.
%
\begin{acknowledgments}
We acknowledge the support from the US National Science Foundation under
Grants No.~PHY-2408989 (R.C.G.) and No.~PHY-2408988 (P.G.K.).
\end{acknowledgments}

\bibliography{Wadati_PT_MAIN}

\end{document}